\documentclass[reprint,amssymb,amsmath,aps,prb,superscriptaddress]{revtex4-1}

\usepackage{graphicx}
\usepackage{amssymb}
\usepackage{epstopdf}
\usepackage{upgreek}
\usepackage{dcolumn}
\usepackage{bm}
\usepackage{gensymb}
\usepackage{braket}
\usepackage{mathtools}
\usepackage{float}
\usepackage{amsmath}
\usepackage{amsfonts}
\usepackage{mathbbol}
\usepackage{color}

\expandafter\ifx\csname package@font\endcsname\relax\else
 \expandafter\expandafter
 \expandafter\usepackage
 \expandafter\expandafter
 \expandafter{\csname package@font\endcsname}%
\fi

\newcommand{\be}{\begin{eqnarray}}
\newcommand{\ee}{\end{eqnarray}}
\newcommand{\bfig}{\begin{figure}}
\newcommand{\efig}{\end{figure}}

\newcommand{\units}[1]{\,\mathrm{#1}}
\begin{document}

\title{Topological phase transition measured in a dissipative metamaterial}
\author{Eric I. Rosenthal}
\thanks{eric.rosenthal@colorado.edu}
\affiliation{JILA, National Institute of Standards and Technology and the University of Colorado, Boulder, Colorado 80309, USA}
\affiliation{Department of Physics, University of Colorado, Boulder, Colorado 80309, USA}
\author{Nicole K. Ehrlich}
\affiliation{JILA, National Institute of Standards and Technology and the University of Colorado, Boulder, Colorado 80309, USA}
\affiliation{Department of Physics, University of Colorado, Boulder, Colorado 80309, USA}
\author{Mark S. Rudner}
\affiliation{Niels Bohr International Academy and Center for Quantum Devices, University of Copenhagen, 2100 Copenhagen, Denmark}
\author{Andrew P. Higginbotham}
\altaffiliation{Microsoft Q, Copenhagen, Denmark}
\affiliation{JILA, National Institute of Standards and Technology and the University of Colorado, Boulder, Colorado 80309, USA}
\affiliation{Department of Physics, University of Colorado, Boulder, Colorado 80309, USA}
\author{K.~W. Lehnert}
\affiliation{JILA, National Institute of Standards and Technology and the University of Colorado, Boulder, Colorado 80309, USA}
\affiliation{Department of Physics, University of Colorado, Boulder, Colorado 80309, USA}
\date{\today}

\begin{abstract}
We construct a metamaterial from radio-frequency harmonic oscillators, and find two topologically distinct phases resulting from dissipation engineered into the system.
These phases are distinguished by a quantized value of bulk energy transport.
The impulse response of our circuit is measured and used to reconstruct the band structure and winding number of circuit eigenfunctions around a dark mode. Our results demonstrate that dissipative topological transport can occur in a wider class of physical systems than considered before. 
\end{abstract}
\maketitle

When a wave propagates through a periodic medium, its properties can be modified dramatically compared to those in free space.
In particular, such ``Bloch waves" may exhibit topological effects: robust behaviors that are insensitive to small changes to the medium.
In solid state systems, the topological characteristics of electronic Bloch waves are responsible for robustly quantized transport in the integer quantum Hall effect \cite{klitzing:1980,thouless:1983}, and the novel characteristics of topological insulators and superconductors \cite{kane:2005, kane:2005b, qi:2011, hasan:2010,bernevig:2013,chiu:2016}.
Importantly, many bosonic and even classical systems, such as cold atoms in optical lattices~\cite{stanescu:2009,goldman:2016}, as well as photonic \cite{rechtsman:2013,lu:2014,mittal:2014,ningyuan:2015,mittal:2016,anderson:2016,cheng:2016,owens:2018,hadad:2018} and mechanical metamaterials \cite{kane:2014,paulose:2015,susstrunk:2016,huber:2016}, can also support waves classified by topological invariants. 


Remarkably, the presence of loss and/or gain can give rise to qualitatively new topological invariants and phenomena~\cite{rudner:2009,diehl:2011,zeuner:2015,rudner:2016,leykam:2017,rakovszky:2017,zhan:2017,weimann:2017}. The topological classification of such non-Hermitian systems is different from that of their lossless, 
Hermitian counterparts. 
In particular, Ref.~\onlinecite{rudner:2009} studied a non-Hermitian quantum problem, and showed that dissipation (particle loss) can lead to a wholly new type of topologically protected quantized transport, which has no analogue in particle conserving systems. Specifically, the mean displacement achieved by a particle between its initialization and decay changes sharply between two quantized values as the relative strengths of the system's two coupling parameters are tuned.

Is this phenomenon specific to first-order-in-time systems governed by an effective non-Hermitian Schr\"{o}dinger equation (e.g., Refs.~\onlinecite{rudner:2009,zeuner:2015,rudner:2016})? Or, is this behavior part of a much more general phenomenology of quantized transport in dissipative wave systems? While it is natural, for example, to construct second-order (harmonic) systems with similar eigenmode winding properties to first-order systems, it is nontrivial that {\it quantized transport} -- which generally depends on both the eigenmode wave functions and the dispersion relation -- could persist in the harmonic case as well. 

\begin{figure}[htb]
\begin{center}
\includegraphics[width=1.0\columnwidth]{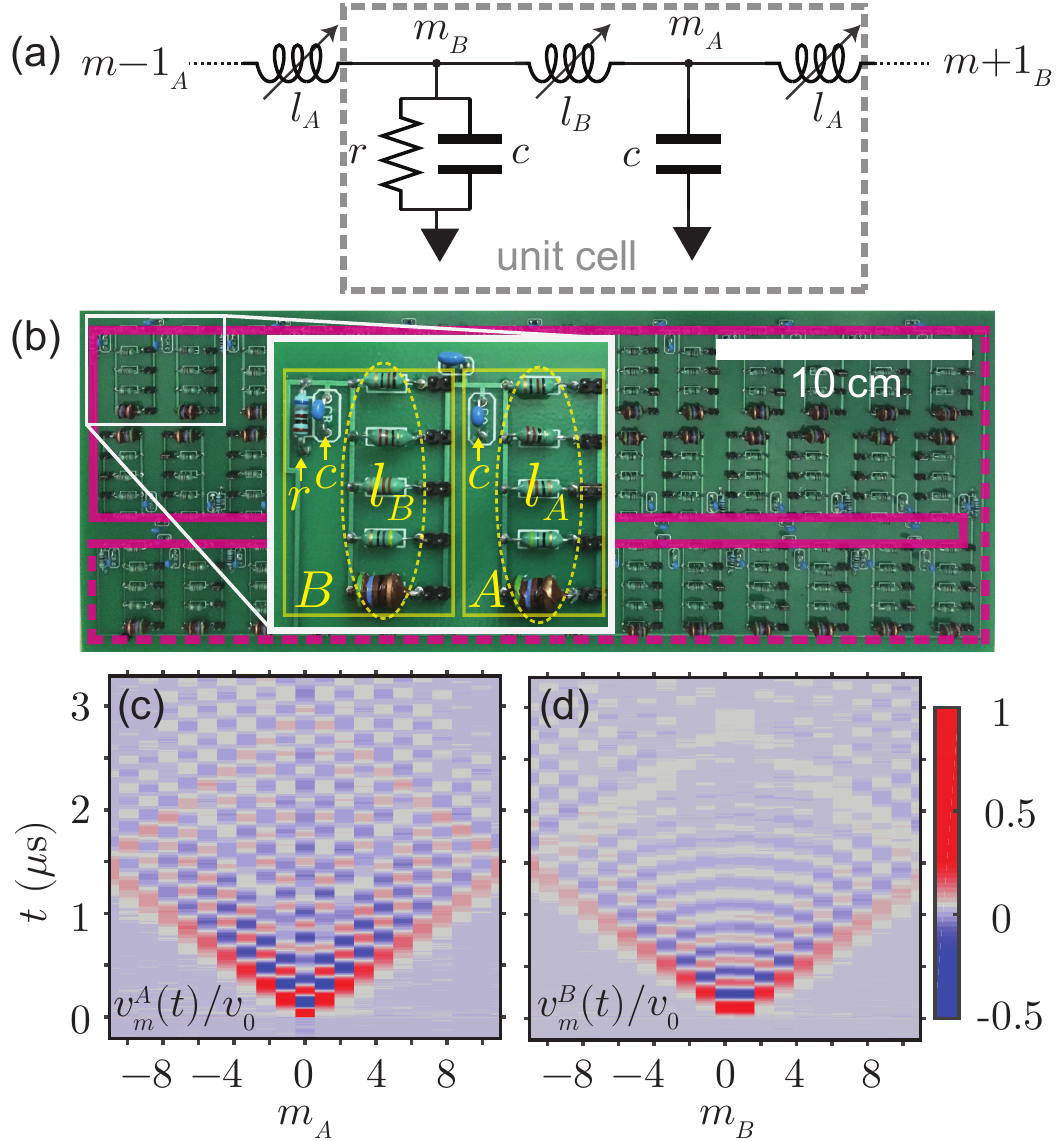}
\caption {(a) A topological metamaterial: circuit nodes, indexed by an integer $m$, are connected to ground with capacitors $c$, and coupled to each other by variable inductors $l_A$ and $l_B$. $B$ sites contain a resistor $r$ to ground, while $A$ sites do not. (b) Circuit photograph. The inset shows one $A$ site and one $B$ site. Coupling inductances are variable by greater than a factor of ten using switch networks. The solid pink line traces a chain of adjacent cells, and the dashed pink line connects the edges of this chain to form periodic boundary conditions. This asymmetric connection has a negligible effect on circuit dynamics; its parasitic capacitance is calculated to be about $3\%$ of the on-site capacitance, and its length (49 $\units{cm}$) is less than about $1\%$ of the wavelength of light at frequencies of interest. (c,d) The time-dependent voltages measured on $A$ and $B$ sites, $v_m^A(t)$ and $v_m^B(t)$, respectively, are plotted as a function of site index $m$ and time $t$ for the case where $l_A = l_B$. Wave forms are normalized to the voltage $v_0$, present at site $m_A = 0$ at time $t = 0$.
}
\label{fig:fig1}
\end{center}
\end{figure}

In this Rapid Communication, we expose the generality of dissipative topological transport by demonstrating its existence in an electromagnetic circuit (Fig.~\ref{fig:fig1}a) governed by {\it harmonic} (non-Schr\"{o}dinger-like) equations of motion. From the impulse response of the circuit we extract the full spatial distribution of dissipated energy. We find that the mean displacement between the location where an energy pulse is injected and where it dissipates takes just one or the other of two quantized values, dependent on the ratio of two coupling parameters (inductances) in the circuit. 
Through analytical arguments and numerical simulations we demonstrate that this transition is {\it topological}, with quantization becoming exact for an ideal realization of the circuit. The impulse response also yields the circuit's eigenmode profiles, which we use to directly image the topological winding characteristic in wave number space that signifies the transition. 
Our observation of quantized transport in a  classical harmonic system opens the way to exploring topological dynamics in a much wider range of physical systems than considered before.

\begin{figure}[htb]
\begin{center}
\includegraphics[width=1.0\linewidth]{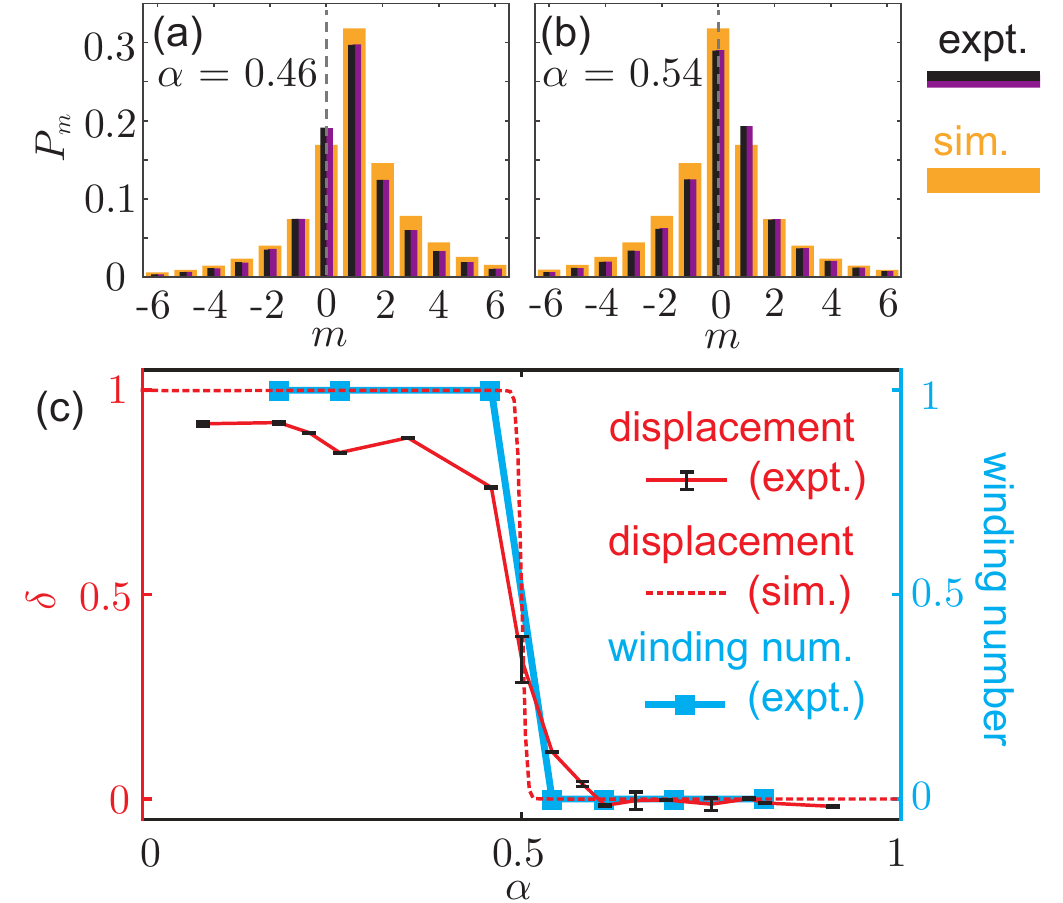}
\caption {Fraction of the total energy dissipated, $P_m$ vs.~$m$, for (a) the topological phase and (b) the trivial phase of our circuit. Gray dashed lines indicate $m = 0$, where an excitation is initialized. Two experimental data sets (black and purple distributions) are compared to simulated distributions (orange) of an `ideal' circuit with no disorder, no additional loss, and 501 unit cells. 
(c) Mean displacement $\delta$ as a function of $\alpha = l_A/(l_B+l_A)$ for the mean of both experimental data sets (solid red line) and a numerical simulation (dashed red line) of an ideal circuit. Black lines show the range of $\delta$ over both experimental runs, which is nearly zero for most values of $\alpha$. A phase transition between $\delta = 1$ and $\delta = 0$ occurs around $\alpha = 0.5$. The behavior of $\delta$ approximates the winding number of $\xi_{j,k}$ around the dark state (blue line).  
}
\label{fig:fig2}
\end{center}
\end{figure}
\textit{Circuit model.---} We investigate a one-dimensional metamaterial with the unit cell defined in Fig.~\ref{fig:fig1}a. Each unit cell, $m$, comprises a lossless `$A$ site' with a capacitor $c$ connected to ground, and a lossy `$B$ site' with a capacitor $c$ and a resistor $r$ connected in parallel to ground. These circuit nodes are connected by tunable coupling inductors $l_A$ and $l_B$.

To study energy transport in both simulation and experiment, we initialize a voltage pulse $v_0$ on one $A$ site, indexed to be $m_0=0$, and measure the impulse response of the circuit by obtaining the time-dependent voltages $v_m^A$ and $v_m^B$ on all $A$ sites and $B$ sites, respectively. The initial pulse contains an energy $\frac12 c v_0^2$. We extract the fraction $P_m$ of this total energy dissipated in each unit cell $m$; the energy displacement $\delta$, characterizing the asymmetric propagation of energy in the circuit, is defined as the mean of the probability distribution, $\{P_m\}$:  
\begin{equation}
\delta = 
\sum_m \left( m - m_0 \right) P_m, \ \ P_m =  \frac{2}{r  c} \int_0^\infty dt\, \left(\frac{v_m^B}{v_0} \right)^2.
\label{Dm}
\end{equation}

The limiting behavior of $\delta$ is easily understood. When $l_B$ approaches infinity while $l_A \ll l_B$ remains finite, a pulse initialized on the lossless $A$ site in unit cell $m$ will dissipate only on the right-adjacent $B$ site, in unit cell $(m+1)$. This yields $\delta = 1$. When $l_A$ approaches infinity while $l_B \ll l_A$ remains finite, the pulse will dissipate only on the left-adjacent $B$ site, within the same unit cell, $m$. This gives $\delta = 0$. Finally, when $l_A = l_B$, the symmetry of the circuit dictates that an excitation will propagate equally to the left and right, so one may expect $\delta = 1/2$ in this case. 

Remarkably, $\delta$ does not smoothly transition between these limiting cases, but rather remains integer-valued as long as $l_A \neq l_B$. To show this, we derive\cite{supp} the equations of motion for this circuit using Kirchhoff's circuit laws. 
In reciprocal space, indexed by the dimensionless wave number $-\pi \leq k < \pi$, dynamics are described in terms of the (complex) variables $v_k^{A(B)}(t)=\sum_m e^{- i k m}\, v_m^{A(B)}(t)$. The circuit equations of motion break into a collection of $2 \times 2$ linear systems, associated with each value of $k$:
\begin{equation}
  \begin{pmatrix}
   \omega_0^2 + \partial_t^2  & g_k \\
    g_k^* & \omega_0^2 + \gamma \partial_t + \partial_t^2
  \end{pmatrix}
  \begin{pmatrix}
    v_k^A \\
    v_k^B
  \end{pmatrix}
  = 0.
\label{EoMk}
\end{equation}
For each $k$, Eq.~(\ref{EoMk}) describes two coupled harmonic oscillators, one of which is damped. The system is parametrized by a resonance frequency $\omega_0 = \sqrt{(l_A+l_B)/(l_A \, l_B \, c)}$, a loss rate $\gamma = 1/(r c)$, and a complex coupling $g_k = - \omega_0^2 \left(\alpha + \beta e^{i k} \right)$ between the $A$ and $B$ sublattice oscillators. 
Here $\alpha = l_A/(l_A+l_B)$ and $\beta = l_B/(l_A+l_B) = 1 - \alpha$. 

When $\alpha = 1/2$, the coupling parameter $g_{k=\pi}$ becomes zero and the sublattices completely decouple. 
At this special point in parameter space the system possesses a ``dark state:'' an excitation on the $A$ sublattice with $k=\pi$ is decoupled from any source of loss and therefore acquires an infinite lifetime~\footnote{In direct space, the dark state corresponds to voltages on neighboring $A$-sites oscillating in anti-phase ($k = \pi$), with zero amplitude on the lossy $B$-sites.}.
Analogous to the situation in non-Hermitian quantum systems~\cite{rudner:2009}, for $\alpha \neq 1/2$ the Bloch eigenfunctions can be classified by a topological winding number that counts the number of times the relative phase of the $A$ and $B$ components runs through $2\pi$ as $k$ traverses the Brillouin zone.

Closely related to this winding number, Eqs.~(\ref{Dm}) and (\ref{EoMk}) predict\cite{supp} the displacement $\delta$ to be quantized as a function of $\alpha$: 
\begin{equation}
\delta = \oint \frac{dk}{2 \pi} \frac{\partial \theta_k}{\partial k} = \left\{ 
	\begin{array}{ll}
		1  & \mbox{if } \alpha < 1/2 \\
		0 & \mbox{if } \alpha > 1/2
	\end{array},
	\right.
\label{DmAnalyticResult}
\end{equation}
where $\theta_k = \units{arg} \{ g_k \}$ is the $k$-dependent angle of $g_k$ in the complex plane.  

Time-domain numerical simulations using a Runge-Kutta method \cite{dormand:1980,shampine:1997} of a system with 501 unit cells and periodic boundary conditions closely match the analytic result of Eq.~(\ref{DmAnalyticResult}). Histograms of the spatial distribution of $P_m$ in the simulation are shown in Fig.~\ref{fig:fig2}a and Fig.~\ref{fig:fig2}b for $\alpha = 0.46$ and $\alpha = 0.54$, respectively. This distribution is computed for many different values of $\alpha$ and is centered around $m=1$ when $\alpha < 0.5$, and $m=0$ when $\alpha > 0.5$. Thus the plot of the simulated $\delta$ vs.~$\alpha$ in Fig.~\ref{fig:fig2}c displays the step-function behavior predicted by Eq.~(\ref{DmAnalyticResult}). 

\begin{figure*}[htb]
\begin{center}
\includegraphics[width=1.0\linewidth]{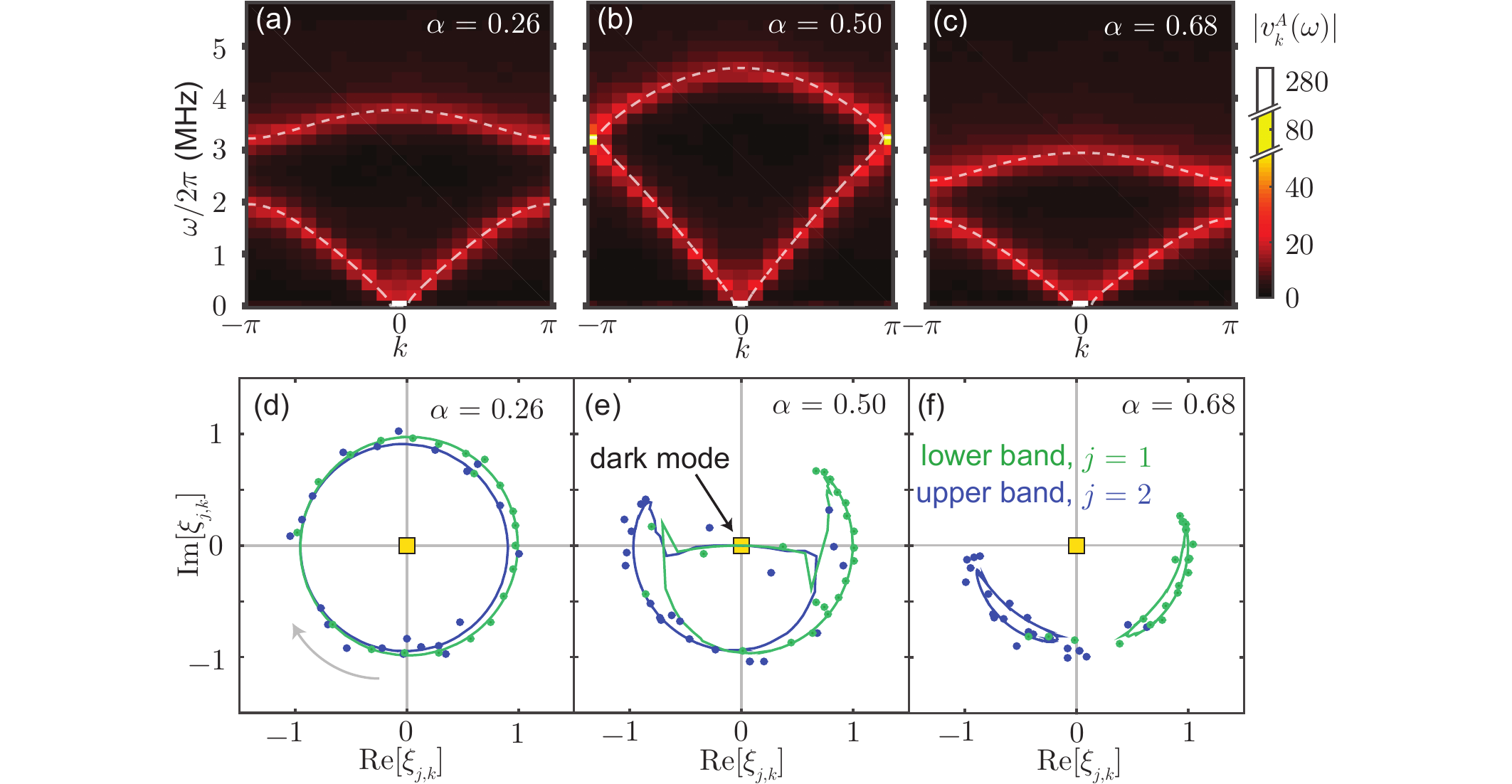}
\caption {(a,b,c) Circuit band structure, exhibited in the Fourier-transformed $A$-sublattice voltages, $|v_k^A(\omega)|$. A dark mode appears at $\alpha = l_A/(l_A+l_B) = 0.5$ and $k = \pi$. The dashed white lines show the real parts of the circuit's complex eigenfrequencies, found by diagonalizing Eq.~(\ref{EoMk}) with the same circuit parameter values that are used in each experiment. Eigenvector sublattice components $\textit{V}_{j,k}^{A}$ and $\textit{V}_{j,k}^{B}$ are extracted from linecuts across the lower ($j=1$) and upper ($j=2$) bands. (d,e,f) The complex ratio $\xi_{j,k} = \textit{V}_{j,k}^{B} / \textit{V}_{j,k}^{A}$, whose winding number around the origin (the location of the dark mode, yellow square) distinguishes the topological phases of the circuit. Experimental points (dots) agree closely with results of a numerical simulation (lines).
}
\label{fig:fig3}
\end{center}
\end{figure*}
\textit{Experimental measurements.---} 
We test the robustness of our theoretical results to real-world imperfections and perturbations with an experiment on a circuit of 21 unit cells and periodic boundary conditions (Fig.~\ref{fig:fig1}b). In the circuit we use capacitors with $c = 381.5 \units{pF}$, resistors with $r = 681 \units{\Omega}$, and coupling inductors with inductances $l_A$ and $l_B$ that can each be set between $5.3 \units{\mu H}$ and $56.7 \units{\mu H}$.
We probe the circuit dynamics by applying a voltage impulse $v_0$ to one $A$ site and measuring the subsequent voltage as a function of time on all circuit nodes using an oscilloscope (Fig.~S1). 

Examples of $A$-site and $B$-site wave forms are plotted in Fig.~\ref{fig:fig1}c and Fig.~\ref{fig:fig1}d, respectively, for the special case when $l_A = l_B$. The longer-lived anti-phase excitations which occur predominantly on the $A$ sublattice are characteristic of the dark state present in this symmetric configuration~\footnote{Note that a finite-size lattice with periodic boundary conditions and an {\it odd} number of unit cells, $N$, does not support a true dark state (Fig.~S2). In this case, the allowed set of crystal momentum values, $k_n = 2  \pi n / N$, where $n = 0, 1, 2, \ldots, N-1$, does not include $k = \pi$. Modes with $k$ values close to $\pi$ still exhibit long lifetimes, with lifetime increasing when $\alpha$ is near 1/2 and the number of unit cells increases.}. Any anti-phase excitations on the $B$ sublattice are heavily damped due to maximum possible coupling to the resistors, the circuit's dominant source of loss. The displacement $\delta$ is calculated from the $B$-site wave forms measured when the circuit has been tuned to different values of $\alpha$ (see, e.g., Fig.~\ref{fig:fig1}d for $\alpha$ = 0.5). 

Although a phase transition is qualitatively apparent in experiment, it is not as sharp as the step function predicted by theory (Eq.~\ref{DmAnalyticResult}) and simulation (Fig.~\ref{fig:fig2}c). Sharpness is reduced by undesired loss, disorder and a finite lattice size (Fig.~S3). A significant source of undesired loss in our experiment is the resistive losses of the coupling inductors, which is between 4 and 29 $\Omega$ (Table~S1), compared to their reactive impedance of between 0.3 $\units{k \Omega}$ to 1.4 $\units{k \Omega}$, respectively, at 4 $\units{MHz}$. 
 
The most striking deviation of our experiment from theory is that the measurements of $\delta$ do not quite approach a value of one as $\alpha$ goes to zero. The finite size of our system combined with periodic boundary conditions contributes to this effect: even at our lowest attainable value of $\alpha$, the impedance of the coupling inductors is small enough to allow some fraction of the initial excitation to propagate more than halfway around the circuit. Such excursions can significantly alter the obtained value of the mean displacement (Fig.~S3). 


To explore the underlying topological nature of the transition observed in Fig.~\ref{fig:fig2}c, we experimentally reconstruct the circuit's band structure and eigenvectors. The impulse response of our circuit encodes all eigenvectors of the family of equations described by Eq.~(\ref{EoMk}), which are all excited simultaneously.
We analyze this response by Fourier transforming the measured wave forms (e.g.,~Figs.~\ref{fig:fig1}c,d) in \textit{both} spatial index and time, in order to obtain the circuit's band structure. The Fourier transform is defined as $v_k^{A(B)}(\omega) = \sum_{t,m} e^{-i (\omega t + k m)} \, v_m^{A(B)}(t)$, where $v_k^A(\omega)$ and $v_k^B(\omega)$ are the complex valued Fourier domain voltages on the $A$ and $B$ sublattices, and $t$ is a discrete time variable with $t = 2 \units{ns} \times \{ 1, 2 \, ... \, 2500 \} $. 

In Figs.~\ref{fig:fig3}a, \ref{fig:fig3}b and \ref{fig:fig3}c we show $|v_k^A(\omega)|$ for $\alpha < 0.5$, $\alpha = 0.5$ and $\alpha > 0.5$, respectively. In all cases, we observe two bands with frequencies that agree closely with the dispersion relation calculated from Eq.~(\ref{EoMk}) (white dashed lines). Note that in the absence of loss, the circuit features a light-like linear dispersion relation at long wavelengths, in contrast to the particle-like quadractic dispersion of the non-Hermitian systems previously shown to exhibit quantized transport~\cite{rudner:2009, zeuner:2015}.

The complex values of $v^A_k(\omega)$ at its intensity peaks encode the $A$-sublattice components of the corresponding Bloch eigenvectors, $\textit{V}_{j,k}^{A}$, where  $j = 1,2$ is a band index labeling the lower and upper bands, respectively. We extract the values of $\textit{V}_{j,k}^{A}$ by taking a vertical linecut of the $A$-sublattice band structure (Fig.~\ref{fig:fig3}a--c) at each value of $k$, and returning the values of $v_k^A(\omega)$ at the two local maxima of each linecut. The corresponding $B$-sublattice component of the eigenvector $\textit{V}_{j,k}^{B}$ is the set of points on linecuts across the $B$-sublattice bands (not shown). 

The $A$ and $B$ eigenvector components can be parameterized by the following $k$-dependent complex ratio:
\begin{equation}
\xi_{j,k} = \textit{V}_{j,k}^{B} / \textit{V}_{j,k}^{A}.
\label{WindingFxn}
\end{equation}
For most values of $k$ and $\alpha$ we see $|\xi_{j,k}| \approx 1$, indicating that the eigenmodes are equally distributed on the two sublattices. 
However near the dark state, i.e., for $\alpha \approx 0.5$ and $k \approx \pi$, $|\xi_{j,k}|$ approaches zero and excitations living primarily on the $A$-sublattice gain long lifetimes (seen by the heavily weighted pixels at $k = \pi$ in Fig.~\ref{fig:fig3}b).

Crucially, the $k$-dependent phase relationship between $\textit{V}_{j,k}^{A}$ and $\textit{V}_{j,k}^{B}$ distinguishes the two topologically distinct phases of the system. Due to the continuity and periodicity of $V_{j,k}^{A}$ and  $V_{j,k}^{B}$  in the Brillouin zone, $\xi_{j,k}$ forms a closed loop in the complex plane. As long as $0 < |\xi_{j,k}| < \infty$, such loops can be classified by the integer number of times that $\xi_{j,k}$ winds around the origin as $k$ goes from 0 to $2\pi$. Thus the topological invariant of our circuit is simply the winding number of $\xi_{j,k}$ around the origin of the complex plane, or, equivalently, around the dark state in parameter space. 

From the experimental data we reconstruct $\xi_{j,k}$ (Figs.~\ref{fig:fig3}d--f). These plots clearly show the winding number transition from 1 to 0 as $\alpha$ is changed from 0.26 to 0.68. These two phases are separated by the case of $\alpha = 0.5$ where $\xi_{j,k}$ intersects the origin and its winding number is undefined. This behavior of the winding number as a function of $\alpha$ is closely approximated by the measured behavior of $\delta$, see Fig.~\ref{fig:fig2}c. 

Unlike many other topological transitions in periodic systems, the transition in our circuit is not associated with the closing of gap in the band structure. Rather, the transition is characterized by a vanishing of the {\it imaginary} part of one of the system's eigenvalues. In particular, the transition persists in a circuit where the $A$-site and $B$-site capacitors are unequal, so that the bandgap does not close for any value of $\alpha$ (Fig.~S4).

{\it Discussion.---} Our observations suggest that dissipative quantized transport may be realized in a much wider class of dynamical systems than previously expected.
Given that high-order linear systems can be reparametrized as first-order systems involving a larger number of variables (bands), it is natural to wonder if the classification for multi-band non-Hermitian quantum walks developed in Ref.~\onlinecite{rudner:2016} can be applied to the second-order system studied in this work.  While the equations of motion for our circuit can be recast as a four-band effective non-Hermitian Schr\"{o}dinger equation [Eq.~(S10)], the physical constraints of the circuit (such as charge conservation) impose a special structure on the resulting Bloch Hamiltonians, which was not accounted for in the arguments of Ref.~\onlinecite{rudner:2016}. 
Elucidating the nature of topological transport in our system, the full range of conditions where it can be realized, and its broader implications (in relation to a general topological classification for dissipative systems\cite{gong:2018}) will be interesting directions for future work. 


\vspace{0.05in}
This work is supported by the Army Research Office under contract W911NF-14-1-0079 and the National Science Foundation under Grant No.
1125844. E.I.R. acknowledges support from the Army Research Office QuaCGR Fellowship. M. R. gratefully acknowledges the support of the European Research Council (ERC) under the European Union Horizon 2020 Research and Innovation Programme (Grant agreement No. 678862), and the Villum Foundation. 

%

\pagebreak
\widetext
\begin{center}
\textbf{\large Supplementary Material for\\ ``Topological phase transition measured in a dissipative metamaterial''}
\end{center}
\setcounter{equation}{0}
\setcounter{figure}{0}
\setcounter{table}{0}
\setcounter{page}{1}
\makeatletter
\renewcommand{\theequation}{S\arabic{equation}}
\renewcommand{\thefigure}{S\arabic{figure}}
\renewcommand{\bibnumfmt}[1]{[S#1]}
\renewcommand{\citenumfont}[1]{S#1}

\section{Circuit equations of motion}
Our circuit equations of motion, Eq.~(2) of the main text, resemble the telegrapher's equations for transmission through a coaxial cable \cite{pozar:2011}, without taking a continuum limit. Here we take a more general model than in the main text, allowing for different capacitances $c_A$ and $c_B$ on the $A$ and $B$ sublattices, respectively. Kirchhoff's laws applied to one unit cell of our circuit (Fig.~1a) yield four equations:
\begin{eqnarray}
v^B_m - v^A_m = l_B \, \frac{d}{dt}{i^B_{m}}, 
\label{KVL2a} \\
i^B_m - i^A_m = c_A \, \frac{d}{dt}{v^A_m}, 
\label{KCL2a} \\
v^A_{m-1} - v^B_m = l_A \, \frac{d}{dt}{i^A_{m-1}}, \label{KVL2b} \\
i^A_{m-1} - i^B_m - \frac{1}{r} v^B_m= c_B \, \frac{d}{dt}{v^B_m}. \label{KCL2b}
\end{eqnarray}
Eq.~(\ref{KVL2a}) describes the time-dependent voltage drop $v_m^{B}(t) - v_m^A(t)$ across the coupling inductor $l_B$ connecting node $m_A$ to node $m_B$, and Eq.~(\ref{KCL2a}) describes the net current flowing into node $m_A$. 
Here $i^B_m(t)$ is the time-dependent current flowing from node $m_B$ to node $m_A$, and $i_m^A(t)$ is the time-dependent current flowing from node $m_A$ to node $(m+1)_B$. Likewise, Eq.~(\ref{KVL2b}) describes the time-dependent voltage drop $v_{m-1}^{A}(t) - v_m^{B}(t)$ across the coupling inductor $l_A$ connecting node $m_B$ to node $(m-1)_A$. The left hand side of Eq.~(\ref{KCL2b}) gives the net current into node $m_B$, where the term $\frac{1}{r}v^B_m$ describes the current flowing to ground through the resistor, $r$.

We combine the four first-order differential equations in Eqs.~(\ref{KVL2a})--(\ref{KCL2b}) into two second-order differential equations:
\begin{eqnarray}
c_A \, \frac{d^2}{dt^2}\phi^A_m + \left(\frac{1}{l_A} + \frac{1}{l_B} \right) \phi^A_m - \frac{1}{l_B} \phi^B_m - \frac{1}{l_A} \phi^B_{m+1} = 0,  \label{EoMa} \\
c_B \, \frac{d^2}{dt^2}\phi^B_m + \frac{1}{r} \frac{d}{dt}\phi^B_m + \left(\frac{1}{l_A} + \frac{1}{l_B} \right) \phi^B_m - \frac{1}{l_A} \phi^A_{m-1} - \frac{1}{l_B} \phi^A_m = 0,
\label{EoMb}
\end{eqnarray}
where we have introduced the ``branch fluxes'' $\phi_m^{A(B)}(t) = \int_{-\infty}^t v_m^{A(B)}(t') \, dt'$. Expressing Eqs.~(\ref{EoMa})~and~(\ref{EoMb}) in Fourier space using $\phi_k^{A(B)}(t) = \sum_m e^{- i k m} \phi_m^{A(B)}(t)$, we find
\begin{equation}
    \left[ \begin{pmatrix}
  \omega_0^2  & g_k \\ 
  g_k^* & \omega_0^2
  \end{pmatrix} + \begin{pmatrix}
    0 & 0 \\
    0 & \gamma
  \end{pmatrix} \partial_t + \begin{pmatrix}
    \eta^{-1} & 0 \\
    0 & \eta
  \end{pmatrix} \partial^2_t \right]   \begin{pmatrix}
    \phi_k^A \\
    \phi_k^B
  \end{pmatrix} = 0,
    \label{DampedHarmonicOsc}
\end{equation}
 where 
\begin{equation}
\eta = \sqrt{c_B / c_A}, \quad \omega_0^2 = \frac{l_A + l_B}{l_A \, l_B \, \sqrt{c_A \, c_B}}, \quad \alpha = \frac{l_A}{l_A + l_B},\ \beta = \frac{l_B}{l_A + l_B}, \quad g_k = -\omega_0^2 \left( \alpha + \beta e^{i k} \right), \quad \gamma = \frac{1}{r \, \sqrt{c_A \, c_B}}.
\end{equation}
Eq.~(\ref{DampedHarmonicOsc}) is trivially differentiated once with respect to time to produce Eq.~(2), using $v_k^{A(B)} = \partial_t \phi_k^{A(B)}$. 

Eqs.~(\ref{KVL2a})---(\ref{KCL2b}) can also be written as four coupled first order equations in a form reminiscent of a non-Hermitian Schr\"{o}dinger equation. To do so, we also write the voltage and current in Fourier space $v_k^{A(B)}(t) = \sum_m e^{-i k m} v_m^{A(B)}(t)$ and $i_k^{A(B)}(t) = \sum_m e^{-i k m} i_m^{A(B)}(t)$. Eqs.~(\ref{KVL2a})---(\ref{KCL2b}) are then combined into a $4 \times 4$ matrix,
\begin{equation}
    \frac{d}{dt}
    \begin{pmatrix}
    i_k^A \\
    i_k^B \\
    v_k^A \\
    v_k^B
    \end{pmatrix} =
    \begin{pmatrix}
    0 & 0 & \frac{1}{l_A} & -\frac{1}{l_A} e^{i k} \\ 
    0 & 0 & -\frac{1}{l_B} & \frac{1}{l_B} \\
    -\frac{1}{c_A} & \frac{1}{c_A} & 0 & 0 \\
    \frac{1}{c_B} e^{-i k} & -\frac{1}{c_B} & 0 & - \frac{1}{r \, c_B}
    \end{pmatrix} 
    \begin{pmatrix}
    i_k^A \\
    i_k^B \\
    v_k^A \\
    v_k^B
    \end{pmatrix}.
    \label{EoM_MatrixForm1}
\end{equation}

Rewriting Eq.~(\ref{EoM_MatrixForm1}) in block form, we obtain:
\begin{equation}
    i \frac{d}{dt} 
    \begin{pmatrix}
    \mathbb{I}_k  \\ 
    \mathbb{V}_k 
    \end{pmatrix} = 
    \begin{pmatrix}
    0 & Q \\ 
    Q^{\dagger} & - i \Gamma 
    \end{pmatrix}
    \begin{pmatrix}
    \mathbb{I}_k  \\ 
    \mathbb{V}_k 
    \end{pmatrix};
    \qquad
    Q = 
    i \omega_0 \begin{pmatrix}
    \sqrt{\beta \, \eta} & - \sqrt{\beta / \eta} \,e^{i k} \\ 
    - \sqrt{\alpha \, \eta} & \sqrt{\alpha / \eta} 
    \end{pmatrix},
    \qquad
    \Gamma = 
    \begin{pmatrix}
    0 & 0 \\ 
    0 & \gamma/\eta 
    \end{pmatrix}.
    \qquad    
    \label{EoM_MatrixForm2}
\end{equation}
Here we introduce the change of variables: $\mathbb{I}_k = \left( \sqrt{\beta} \, l_A \, i_k^A\  \sqrt{\alpha} \, l_B \, i_k^B \right)^T$ and $\mathbb{V}_k = \left( 1/(\sqrt{\eta} \, \omega_0) \, v_k^A\ \sqrt{\eta} / \omega_0 \, v_k^B \right)^T$. The two real, second-order-in-time equations of motion (Eq.~\ref{DampedHarmonicOsc}) have now been reformulated as four complex, first-order-in-time equations. The effective non-Hermitian Hamiltonian appearing in Eq.~(S10) has a special structure, involving many zeros, due to the physical nature of the original circuit. Such structure is not accounted for in existing topological classifications of non-Hermitian systems (see the main text).

\section{Derivation of quantized energy transport} 
We start with Eq.~(1) of the main text, which defines the mean energy displacement $\delta$.
Using $m v_m^B(t) = - i \oint \frac{dk}{2 \pi} \left( \frac{d}{dk} e^{i k m} \right) v_k^B(t)$, we obtain
\begin{equation}
\delta = \frac{i}{r \, \epsilon_0} \int_0^{\infty} dt \oint \frac{dk}{2 \pi} v_k^{B*} \frac{\partial v_k^B}{\partial k},
\label{DmIntegralForm}
\end{equation}
where $\epsilon_0 = \frac{1}{2} c_A v_0^2$ is the total energy in the circuit at time $t=0$. We assume a lattice of infinite size so that $-\pi \leq k < \pi$ is a continuous variable, although the unit cell index $-\infty < m < \infty$ remains a discrete variable.

We now analytically show that $\delta$ is quantized, with a value dependent on $\alpha$. 
First we define the two component vector $\vec{v}_k = (v_k^A\ v_k^B)^T$, and the matrix operator $M_k$ appearing in Eq.~(2) of the main text, such that $\partial_t \vec{v}_k = M_k\vec{v}_k$. Next we apply a unitary transformation to eliminate the complex phase of the coupling $g_k = -\omega_0^2 ( \alpha + \beta e^{ik} )$, and obtain purely real equations of motion (in the rotated frame) for each $k$.
Let $\tilde{\vec{v}}_k = U_k\vec{v}_k$, where $U_k = e^{i (1 - \hat{\sigma}_z) \theta_k / 2}$. 
Here $\sigma_z$ is the third Pauli matrix.
The unitary transformation shifts the phase of the $v_k^B$ component of $\vec{v}_k$ by the angle $\theta_k$: $\tilde{\vec{v}}_k = (v_k^A\ v_k^B e^{i\theta_k})^T$.
Choosing $\theta_k = \units{arg} \{ g_k \} $, we obtain the transformed equation of motion $\partial_t \tilde{\vec{v}}_k = \tilde{M}_k\tilde{\vec{v}}_k$, with
\begin{equation}
    \tilde{M}_k = U_kM_kU_k^\dagger = \begin{pmatrix}
    \omega_0^2 + \eta^{-1} \partial_t^2 & |g_k| \\ 
    |g_k| & \omega_0^2 + \gamma \partial_t + \eta \partial_t^2
    \end{pmatrix}.
    \label{UnitaryTransformation}
\end{equation}
Importantly, the (real-valued) initial condition $\vec{v}_k(t=0) = (v_0\ 0)^T$ is invariant under the unitary transformation. Therefore, since the matrix $\tilde{M}$ in Eq.~(\ref{UnitaryTransformation}) is real, the solutions $\tilde{\vec{v}}_k(t)$ are real-valued for all $t$.

We now return to Eq.~(\ref{DmIntegralForm}) to compute the expected energy displacement, $\langle m\rangle$.
Inverting the unitary transformation above, we observe that $v_k^B(t) = \tilde{v}_k e^{-i\theta_k}$, with $\tilde{v}_k$ real. Substituting into Eq.~(\ref{DmIntegralForm}) gives
\begin{equation}
\delta = \frac{i}{r \, \epsilon_0} \int_0^{\infty} dt \oint \frac{dk}{2 \pi} \left[ \tilde{v}_k \frac{\partial \tilde{v}_k}{\partial k} - i \tilde{v}_k^2 \frac{\partial \theta_k}{\partial k}\right].
\label{DmIntegralForm2}
\end{equation}
The first term inside the integral in Eq.~(\ref{DmIntegralForm2}) is zero, because $\tilde{v}_k \, \partial_k \tilde{v}_k$ is a total derivative and its integral vanishes when evaluated over a closed contour. 
Next, we define the time-dependent total energy stored in modes with wavenumber $k$ as $\epsilon_k(t) = \frac{1}{2} c_A \, |v_k^A(t)|^2 \, + \, c_B \, \frac{1}{2} |v_k^B(t)|^2 + \frac{1}{2} l_A \, |i_k^A(t)|^2 + \frac{1}{2} \, l_B |i_k^B(t)|^2$. 
The rate of change of $\epsilon_k(t)$ is proportional to the power dissipated in the $B$-site resistors: $\partial_t \epsilon_k = - \tilde{v}_k^2 / r$.
Using this relation, Eq.~(\ref{DmIntegralForm2}) becomes
\begin{equation}
\delta = - \frac{1}{\epsilon_0} \oint \frac{dk}{2 \pi} \frac{\partial \theta_k}{\partial k} \int_{0}^{\infty} dt \frac{\partial \epsilon_k}{\partial t}.
\label{DmIntegralForm3}
\end{equation}
Here we used the fact that $\theta_k$ is independent of time to move it ahead of the time-integral. We evaluate the time-dependent integral in Eq.~(\ref{DmIntegralForm3}) to get:

\begin{equation}
\delta = \oint \frac{dk}{2 \pi} \frac{\partial \theta_k}{\partial k} = 
    \begin{cases}
		1  & \mbox{if } \alpha < 1/2 \\
		0 & \mbox{if } \alpha > 1/2    
    \end{cases},
\label{DmIntegralFormFinal}
\end{equation}

The energy displacement $\delta$ is therefore equal to the winding number of $\units{arg} \{ g_k \}$, as $k$ is taken around the first Brillouin zone.

\section{Experimental setup and circuit non-idealities}
\begin{figure}[tb]
\begin{center}
\includegraphics[width=1.0\linewidth]{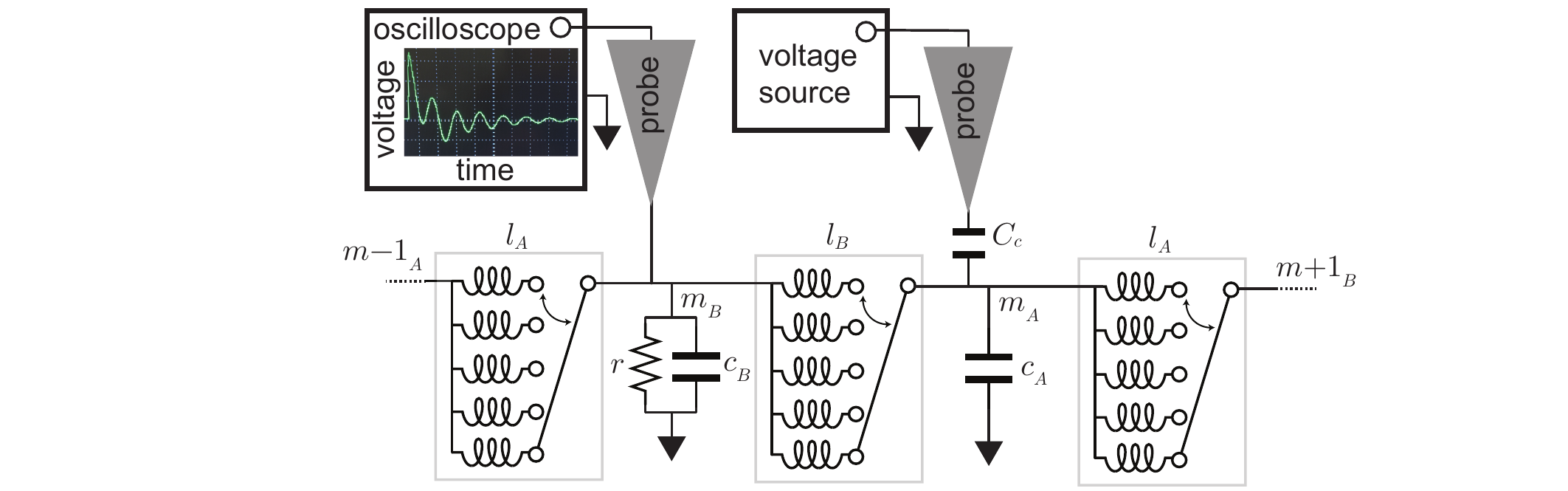}
\caption {\small Experimental schematic: coupling inductors are discretely tunable by switching between five different inductors. The circuit is excited by a near-instantaneous voltage pulse at one $A$-site. Voltage as a function of time is then measured at all nodes of the circuit. 
}
\label{fig:supp1}
\end{center}
\end{figure}

A full experimental schematic is shown in Fig.~\ref{fig:supp1}. A coupling capacitor $C_c = 3 \pm 0.25 \units{pF}$ is placed on each $A$-site in order to initialize a voltage pulse with a rise time of approximately 34 $\units{ns}$, significantly faster than other circuit dynamics. Voltage as a function of time on all circuit nodes is measured using an oscilloscope, with voltage on each node measured in a separate experimental run of the circuit. 

Every node of our circuit has a $c = 381.5 \pm 1.5 \units{pF}$ capacitor connected to ground, and alternating circuit nodes (the $B$-sites) each have a resistor $r = 681 \pm 3 \units{\Omega}$ connected to ground in parallel with the capacitor. The coupling inductors ($l_A$ or $l_B$) are discretely tunable using a switch network composed of five different inductors (Table~\ref{table:table1}). The inductor combinations used for each of the 16 values of $\alpha$ measured in experiment are shown in Table~\ref{table:table2}. 

\begin{table}[b]
\begin{center}
\begin{tabular}{ | m{2.8cm} || m{1.9cm} | m{1.9cm} | }
\hline
Inductance ($\units{\mu H}$) & ESR ($\units{\Omega}$) & SRF (MHz) \\ 
\hline\hline
12.6 $\pm$ 0.2 & 4.3 & 16\\
\hline
22.7 $\pm$ 0.2 & 8.2 & 13\\
\hline 
35.1 $\pm$ 0.3 & 15 & 10\\ 
\hline
48.5 $\pm$ 0.3 & 20 & 8.5\\
\hline
56.7 $\pm$ 0.2 & 29 & 6.8\\
\hline 
\end{tabular}
\caption{Inductor equivalent series resistance (ESR) measured at 4 MHz, and self-resonance frequency (SRF). The SRF of all inductors is above the maximum circuit eigenfrequency of about 4.5 MHz (seen in Fig.~3b). To create the maximum and minimum values of $\alpha$ measured in experiment, we connect all five inductors in parallel to create an effective inductance of $l_A = 5.3 \units{nH}$. }
\label{table:table1}
\end{center}
\end{table}

\begin{table}[htb]
\begin{center}
\begin{tabular}{ | m{1cm} || m{1.5cm} | m{1.5cm} | m{2cm} | }
\hline
$\alpha$ & $l_A \, (\units{\mu H})$ & $l_B \, (\units{\mu H})$ & $\omega_0 / 2 \pi \, (\units{MHz})$ \\ 
\hline\hline
0.09 & 5.3 & 56.7 & 3.70 \\
\hline
0.18 & 12.6 & 56.7 & 2.54 \\
\hline 
0.22 & 12.6 & 48.5 & 2.58 \\ 
\hline
0.26 & 12.6 & 35.1 & 2.68 \\
\hline
0.36 & 12.6 & 22.7 & 2.86 \\
\hline 
0.46 & 48.5 & 56.7 & 1.59 \\
\hline 
0.50 & 12.6 & 12.6 & 3.25 \\
\hline 
0.54 & 56.7 & 48.5 & 1.59 \\
\hline 
0.58 & 48.5 & 35.1 & 1.80 \\
\hline 
0.62 & 56.7 & 35.1 & 1.75 \\
\hline 
0.64 & 22.7 & 12.6 & 2.86 \\
\hline 
0.68 & 48.5 & 22.7 & 2.07 \\
\hline 
0.74 & 35.1 & 12.6 & 2.68 \\
\hline 
0.79 & 48.5 & 12.6 & 2.58 \\
\hline 
0.82 & 56.7 & 12.6 & 2.54 \\
\hline 
0.91 & 56.7 & 5.3 & 3.70 \\
\hline 
\end{tabular}
\caption{The coupling inductor combinations used to tune the topological circuit to the 16 different values of $\alpha$ discussed in this work.}
\label{table:table2}
\end{center}
\end{table}

\begin{figure}[htb]
\begin{center}
\includegraphics[width=1.0\linewidth]{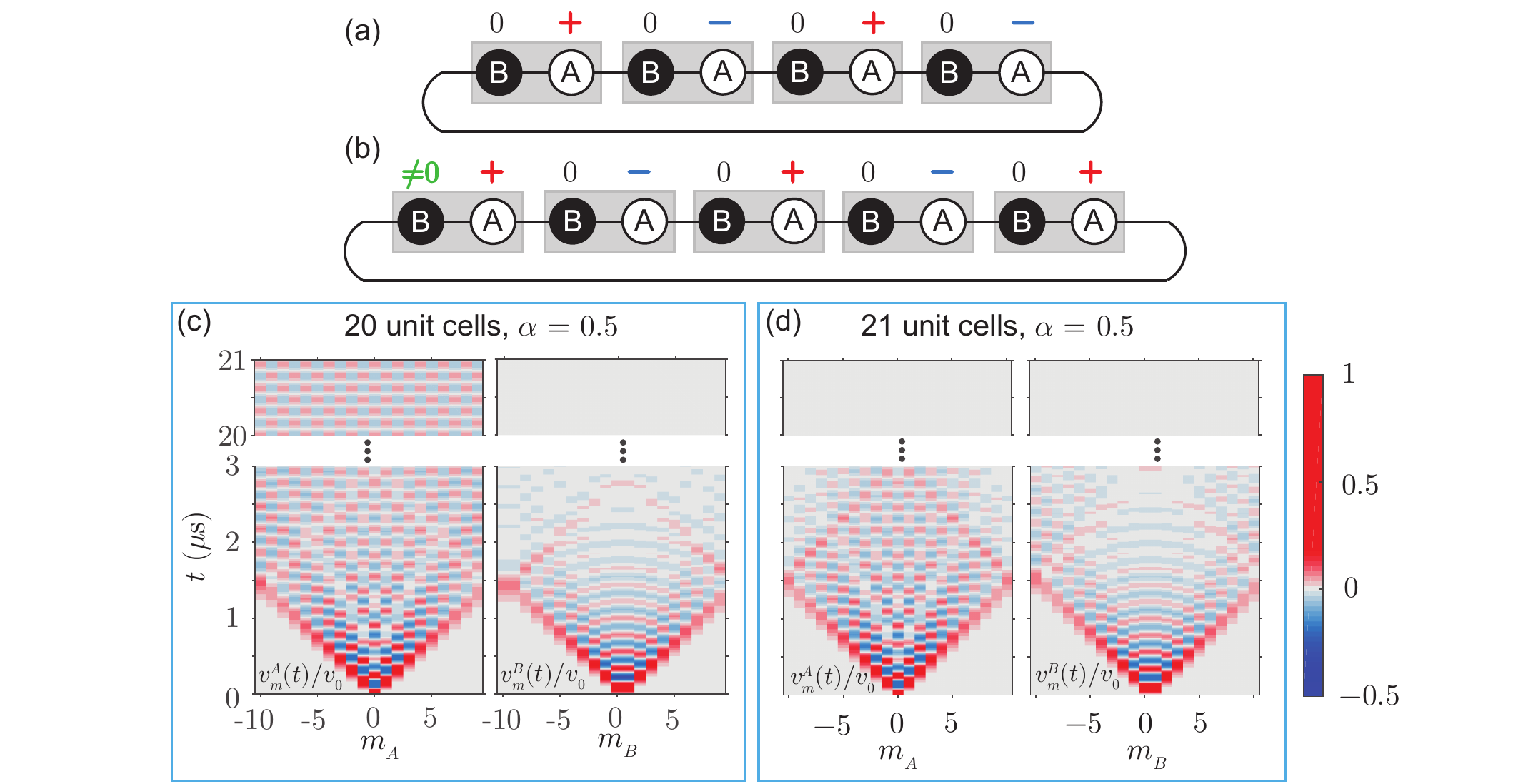}
\caption {(a) A finite-sized topological circuit with periodic boundary conditions and an even number of unit cells. The dark state has out-of-phase antinodes on all $A$-sites (represented by red ``$+$" and blue ``$-$" signs) and nodes on all lossy $B$-sites, so it is decoupled from any source of loss. (b) A finite-sized circuit with periodic boundary conditions and an odd number of unit cells. The dark state cannot fully exist; in this case, the voltages on the leftmost and rightmost $A$-sites have the same sign, so the voltage on the $B$-site between them will not be zero at all times. (c) Simulations of a lattice with an even number of unit cells. The leftmost panel shows voltage as a function of time on all $A$-sites, where a dark state persists as time goes to infinity. (d) In contrast, voltage on all $A$-sites eventually decays to zero in a system with an odd number of cells. This simulation reproduces the same behavior seen in our experiment (Fig.~1c,d).
}
\label{fig:supp2}
\end{center}
\end{figure}

\begin{figure}[htb]
\begin{center}
\includegraphics[width=1.0\linewidth]{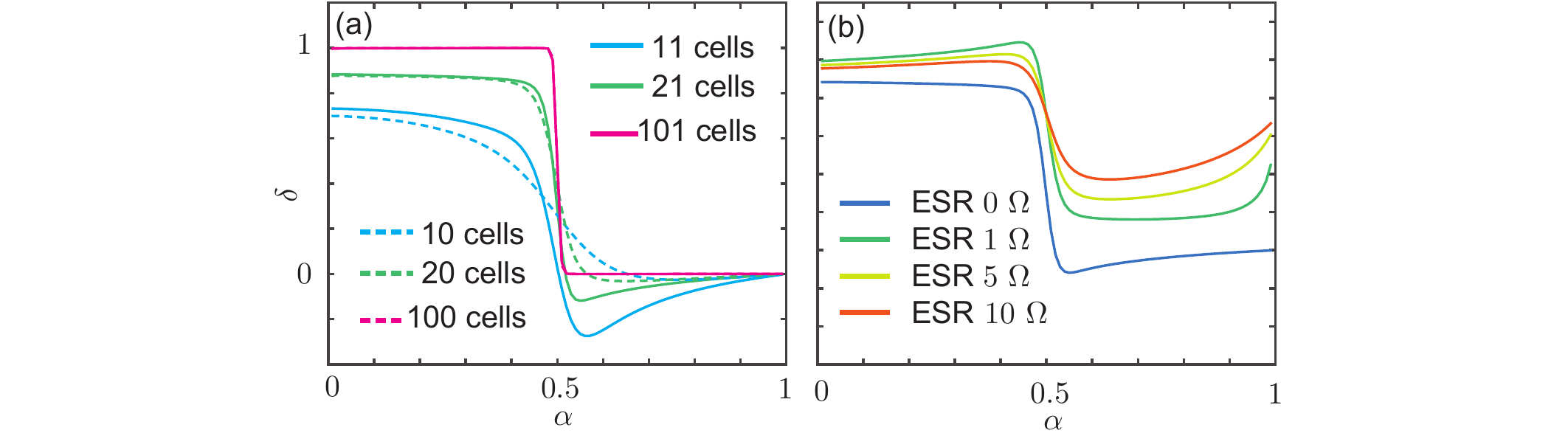}
\caption {Simulated topological transitions: for all simulations shown, a numerical model of the circuit with periodic boundary conditions is simulated in the time and spatial domains (Eqs.~(S1)---(S4)). In each model circuit, $c$ and $r$ are equal to their mean value in experiment, and $\alpha$ is parameterized by sweeping $l_A$ from 0 to 50 $\units{\mu H}$ while $l_B$ is commensurately swept from 50 to 0 $\units{\mu H}$. (a) Lattice size is varied in a circuit with no coupling loss or disorder. (b) Equivalent series resistance (ESR) of the coupling inductors is varied in a circuit with 21 unit cells and no disorder. 
}
\label{fig:supp3}
\end{center}
\end{figure}

The actual device employed in our experiment (Fig.~1b) is of course not a perfect realization of the idealized circuit drawn in Fig.~1a. As a result, various factors reduce the sharpness of the observed transition of $\delta$ as a function of $\alpha$, as compared with the ideal case:
\begin{itemize}
    \item \textit{Finite lattice size:} our circuit has periodic boundary conditions and contains 21 unit cells. A true dark state cannot actually exist in a lattice with periodic boundary conditions and an \textit{odd} number of unit cells (Fig.~\ref{fig:supp2}). Even for an ideal system with an even number of unit cells, the plot of $\delta$ vs. $\alpha$ becomes significantly rounded for systems of this size. Simulations in Fig.~\ref{fig:supp3}a show that $\delta = 0.88$ at $\alpha = 0.01$ for a lattice of 20 unit cells, and $c$ and $r$ equal to their mean experimental values. For comparison, in the experiment we measure $\delta = 0.92$ when $\alpha = 0.08$. 

    \item \textit{Undesired loss:} Table~\ref{table:table1} lists the equivalent series resistances (ESR) of the coupling inductors, the dominant source of undesired loss in our circuit. ESR values are between 1.3\% to 2.0\% of the reactive impedance of the coupling inductors at 4 $\units{MHz}$. Simulations of circuits with different amounts of coupling inductor ESR are compared in Fig.~\ref{fig:supp3}b.       
    
    \item \textit{Disorder:} All inductor, resistor and capacitor values differ from their mean by $ \pm$1.5$\%$ or less. However, simulations (not shown) indicate that disorder has a negligible effect on transition sharpness compared to finite-size effects and extra loss.
\end{itemize}
Additionally, note that the resonance frequency $\omega_0$ changes between $1.6$ and $3.7 \units{MHz}$ for the different values of $\alpha$ at which the circuit is measured at (Table~\ref{table:table2}). This distribution contributes to the scatter of the data points shown in Fig.~2c; finite-size effects, for instance, are highly dependent on the time-scale of the circuit dynamics.

Finally, a direct-current voltage offset of about 5 mV (the voltage difference between the ground of the voltage source and the ground of the oscilloscope) is also present in our data. We modify our reconstruction of $P_m$ to remove this effect, by using the time average $\frac{1}{T} \sum_t \Delta t \, v^B_m(t) $ to estimate the offset. Here, $\Delta t = 2 \units{ns}$ (the time-resolution of the oscilloscope) and $T = 4.86 \units{\mu s}$ is the time-span of the measured traces, beginning from the maximum of the voltage excitation.
Therefore,
\begin{equation}
P_m = 2\gamma \left[\sum_t \Delta t \left(\frac{v_m^B(t)}{v_0} \right)^2 - \frac{1}{T} \left( \sum_t \Delta t \frac{v_m^B(t)}{v_0} \right)^2 \right],
\label{PmSupplement}
\end{equation}
where $v_0 = 158 \units{mV}$ is the initial voltage applied to the circuit. Eq.~(\ref{PmSupplement}) is used to generate the experimental plot of $\delta$ vs. $\alpha$ shown in Fig.~2c.

\section{Topological phase transition in an asymmetric circuit}
Unlike many other topological transitions in periodic systems, the transition in our circuit is not associated with the closing of gap in the band structure. Rather, the transition is characterized by a vanishing of the {\it imaginary} part of one of the system's eigenvalues. In particular, the transition persists in a circuit where the $A$-site and $B$-site capacitors are unequal, so that the bandgap does not close for any value of $\alpha$ (Fig.~S4).

\begin{figure}[htb]
\begin{center}
\includegraphics[width=1.0\linewidth]{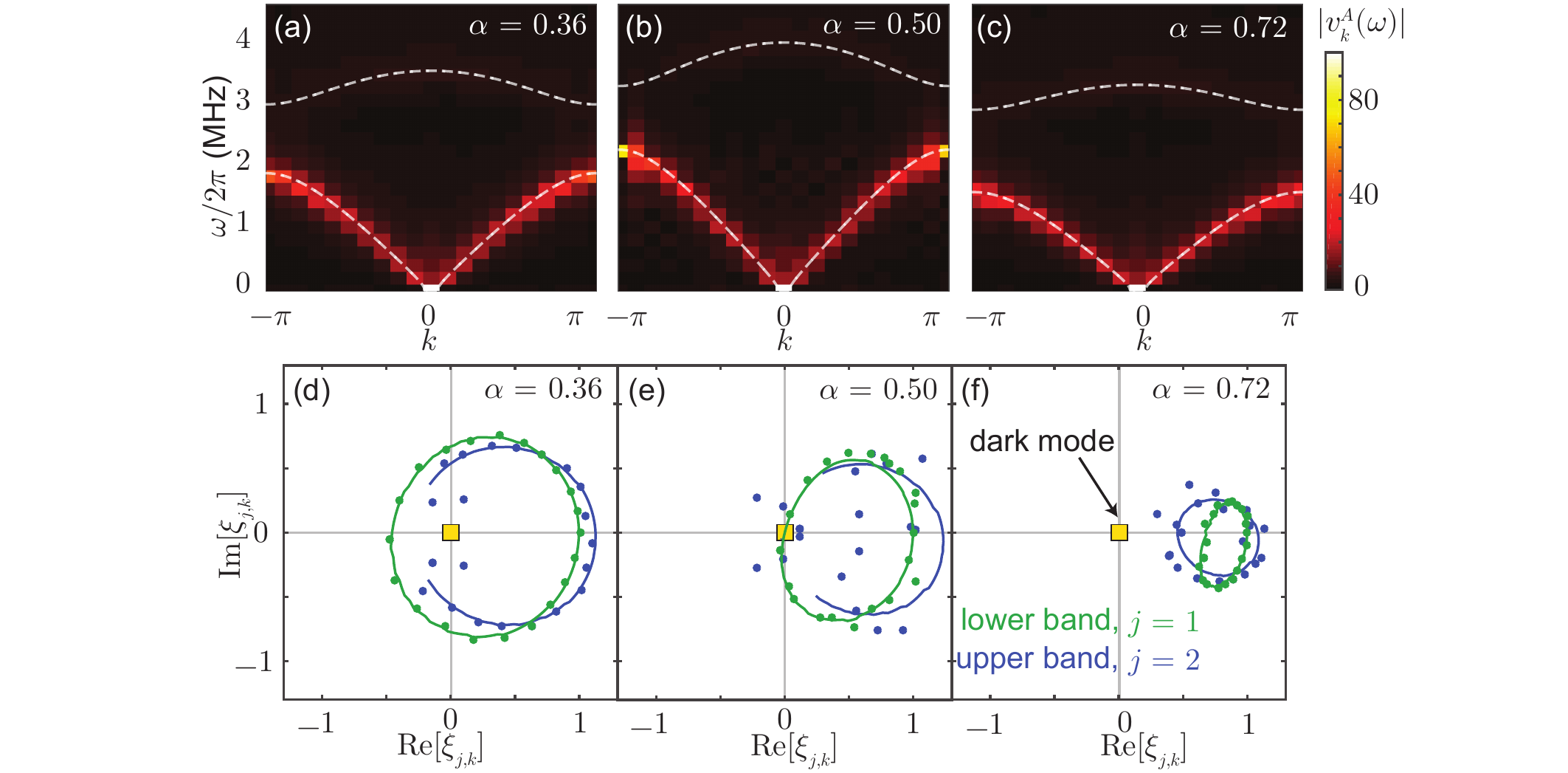}
\caption {(a,b,c) Measured $A$-sublattice band structure $|v_k^A(\omega)|$ of a circuit where $c_A \neq c_B$. Eigenvector sublattice components $\textit{V}_{j,k}^{A}$ and $\textit{V}_{j,k}^{B}$ correspond to linecuts along the lower and upper bands ($j=1$ and $j=2$, respectively). (d,e,f) The complex ratio $\xi_{j,k} = \textit{V}_{j,k}^{B} / \textit{V}_{j,k}^{A}$, whose winding number over the first Brillouin zone distinguishes the topological phase of the circuit. Experimental points (dots) agree with results of a numerical simulation (lines), which are not plotted for all values of $k$ in (d) and (e) due to the weak amplitude of the upper band eigenvector sublattice components near $k=\pi$.
}
\label{fig:supp4}
\end{center}
\end{figure}

We also construct a circuit with unequal capacitors on the $A$-sites and $B$-sites in order to demonstrate that the topological phase transition is robust to the breaking of this symmetry. All parameters in this circuit are chosen to be the same as the circuit discussed in the main text, except for the $A$-site capacitors which are $815.5 \pm 1.5 \units{pF}$. The band structure and winding function $\xi_{j,k}$ of the asymmetric circuit are shown in Fig.~\ref{fig:supp4} for three different values of $\alpha$. The experimental band structures (color plots in Fig.~\ref{fig:supp4}) agree with the real parts of the dispersion relation calculated by diagonalizing Eq.~(\ref{DampedHarmonicOsc}) (white dashed line). 

Unlike the symmetric circuit discussed in the main text, upper band excitations of the asymmetric circuit are significantly weaker than those in the lower band. This effect is expected by diagonalizing Eq.~(\ref{DampedHarmonicOsc}): when $c_A \neq c_B$, upper and lower band solutions will have different imaginary components across all values of $k$ (this is not true when $c_A = c_B$). In our case of $c_A > c_B$, the upper band solution to $\omega(k)$ has a larger imaginary value than the lower band indicating that excitations in the upper band are damped more heavily.
Despite this asymmetry, the winding function $\xi_{j,k}$ of this circuit still clearly encircles the origin when $\alpha < 0. 5$ and does not when $\alpha > 0.5$. 

\vspace{0.1in}

[S1] D. M. Pozar, \textit{Microwave Engineering. 4th} (2011).


\begin{thebibliography}{36}%
\makeatletter
\providecommand \@ifxundefined [1]{%
 \@ifx{#1\undefined}
}%
\providecommand \@ifnum [1]{%
 \ifnum #1\expandafter \@firstoftwo
 \else \expandafter \@secondoftwo
 \fi
}%
\providecommand \@ifx [1]{%
 \ifx #1\expandafter \@firstoftwo
 \else \expandafter \@secondoftwo
 \fi
}%
\providecommand \natexlab [1]{#1}%
\providecommand \enquote  [1]{``#1''}%
\providecommand \bibnamefont  [1]{#1}%
\providecommand \bibfnamefont [1]{#1}%
\providecommand \citenamefont [1]{#1}%
\providecommand \href@noop [0]{\@secondoftwo}%
\providecommand \href [0]{\begingroup \@sanitize@url \@href}%
\providecommand \@href[1]{\@@startlink{#1}\@@href}%
\providecommand \@@href[1]{\endgroup#1\@@endlink}%
\providecommand \@sanitize@url [0]{\catcode `\\12\catcode `\$12\catcode
  `\&12\catcode `\#12\catcode `\^12\catcode `\_12\catcode `\%12\relax}%
\providecommand \@@startlink[1]{}%
\providecommand \@@endlink[0]{}%
\providecommand \url  [0]{\begingroup\@sanitize@url \@url }%
\providecommand \@url [1]{\endgroup\@href {#1}{\urlprefix }}%
\providecommand \urlprefix  [0]{URL }%
\providecommand \Eprint [0]{\href }%
\providecommand \doibase [0]{http://dx.doi.org/}%
\providecommand \selectlanguage [0]{\@gobble}%
\providecommand \bibinfo  [0]{\@secondoftwo}%
\providecommand \bibfield  [0]{\@secondoftwo}%
\providecommand \translation [1]{[#1]}%
\providecommand \BibitemOpen [0]{}%
\providecommand \bibitemStop [0]{}%
\providecommand \bibitemNoStop [0]{.\EOS\space}%
\providecommand \EOS [0]{\spacefactor3000\relax}%
\providecommand \BibitemShut  [1]{\csname bibitem#1\endcsname}%
\let\auto@bib@innerbib\@empty
\bibitem [{\citenamefont {Klitzing}\ \emph {et~al.}(1980)\citenamefont
  {Klitzing}, \citenamefont {Dorda},\ and\ \citenamefont
  {Pepper}}]{klitzing:1980}%
  \BibitemOpen
  \bibfield  {author} {\bibinfo {author} {\bibfnamefont {K.~v.}\ \bibnamefont
  {Klitzing}}, \bibinfo {author} {\bibfnamefont {G.}~\bibnamefont {Dorda}}, \
  and\ \bibinfo {author} {\bibfnamefont {M.}~\bibnamefont {Pepper}},\ }\href
  {\doibase 10.1103/PhysRevLett.45.494} {\bibfield  {journal} {\bibinfo
  {journal} {Phys. Rev. Lett.}\ }\textbf {\bibinfo {volume} {45}},\ \bibinfo
  {pages} {494} (\bibinfo {year} {1980})}\BibitemShut {NoStop}%
\bibitem [{\citenamefont {Thouless}(1983)}]{thouless:1983}%
  \BibitemOpen
  \bibfield  {author} {\bibinfo {author} {\bibfnamefont {D.~J.}\ \bibnamefont
  {Thouless}},\ }\href {\doibase 10.1103/PhysRevB.27.6083} {\bibfield
  {journal} {\bibinfo  {journal} {Phys. Rev. B}\ }\textbf {\bibinfo {volume}
  {27}},\ \bibinfo {pages} {6083} (\bibinfo {year} {1983})}\BibitemShut
  {NoStop}%
\bibitem [{\citenamefont {Kane}\ and\ \citenamefont
  {Mele}(2005{\natexlab{a}})}]{kane:2005}%
  \BibitemOpen
  \bibfield  {author} {\bibinfo {author} {\bibfnamefont {C.~L.}\ \bibnamefont
  {Kane}}\ and\ \bibinfo {author} {\bibfnamefont {E.~J.}\ \bibnamefont
  {Mele}},\ }\href {\doibase 10.1103/PhysRevLett.95.146802} {\bibfield
  {journal} {\bibinfo  {journal} {Phys. Rev. Lett.}\ }\textbf {\bibinfo
  {volume} {95}},\ \bibinfo {pages} {146802} (\bibinfo {year}
  {2005}{\natexlab{a}})}\BibitemShut {NoStop}%
\bibitem [{\citenamefont {Kane}\ and\ \citenamefont
  {Mele}(2005{\natexlab{b}})}]{kane:2005b}%
  \BibitemOpen
  \bibfield  {author} {\bibinfo {author} {\bibfnamefont {C.~L.}\ \bibnamefont
  {Kane}}\ and\ \bibinfo {author} {\bibfnamefont {E.~J.}\ \bibnamefont
  {Mele}},\ }\href {\doibase 10.1103/PhysRevLett.95.226801} {\bibfield
  {journal} {\bibinfo  {journal} {Phys. Rev. Lett.}\ }\textbf {\bibinfo
  {volume} {95}},\ \bibinfo {pages} {226801} (\bibinfo {year}
  {2005}{\natexlab{b}})}\BibitemShut {NoStop}%
\bibitem [{\citenamefont {Qi}\ and\ \citenamefont {Zhang}(2011)}]{qi:2011}%
  \BibitemOpen
  \bibfield  {author} {\bibinfo {author} {\bibfnamefont {X.-L.}\ \bibnamefont
  {Qi}}\ and\ \bibinfo {author} {\bibfnamefont {S.-C.}\ \bibnamefont {Zhang}},\
  }\href {\doibase 10.1103/RevModPhys.83.1057} {\bibfield  {journal} {\bibinfo
  {journal} {Rev. Mod. Phys.}\ }\textbf {\bibinfo {volume} {83}},\ \bibinfo
  {pages} {1057} (\bibinfo {year} {2011})}\BibitemShut {NoStop}%
\bibitem [{\citenamefont {Hasan}\ and\ \citenamefont
  {Kane}(2010)}]{hasan:2010}%
  \BibitemOpen
  \bibfield  {author} {\bibinfo {author} {\bibfnamefont {M.~Z.}\ \bibnamefont
  {Hasan}}\ and\ \bibinfo {author} {\bibfnamefont {C.~L.}\ \bibnamefont
  {Kane}},\ }\href {\doibase 10.1103/RevModPhys.82.3045} {\bibfield  {journal}
  {\bibinfo  {journal} {Rev. Mod. Phys.}\ }\textbf {\bibinfo {volume} {82}},\
  \bibinfo {pages} {3045} (\bibinfo {year} {2010})}\BibitemShut {NoStop}%
\bibitem [{\citenamefont {Bernevig}(2013)}]{bernevig:2013}%
  \BibitemOpen
  \bibfield  {author} {\bibinfo {author} {\bibfnamefont {B.~A.}\ \bibnamefont
  {Bernevig}},\ }\href@noop {} {\enquote {\bibinfo {title} {Topological
  insulators and topological superconductors},}\ } (\bibinfo {year}
  {2013})\BibitemShut {NoStop}%
\bibitem [{\citenamefont {Chiu}\ \emph {et~al.}(2016)\citenamefont {Chiu},
  \citenamefont {Teo}, \citenamefont {Schnyder},\ and\ \citenamefont
  {Ryu}}]{chiu:2016}%
  \BibitemOpen
  \bibfield  {author} {\bibinfo {author} {\bibfnamefont {C.-K.}\ \bibnamefont
  {Chiu}}, \bibinfo {author} {\bibfnamefont {J.~C.~Y.}\ \bibnamefont {Teo}},
  \bibinfo {author} {\bibfnamefont {A.~P.}\ \bibnamefont {Schnyder}}, \ and\
  \bibinfo {author} {\bibfnamefont {S.}~\bibnamefont {Ryu}},\ }\href {\doibase
  10.1103/RevModPhys.88.035005} {\bibfield  {journal} {\bibinfo  {journal}
  {Rev. Mod. Phys.}\ }\textbf {\bibinfo {volume} {88}},\ \bibinfo {pages}
  {035005} (\bibinfo {year} {2016})}\BibitemShut {NoStop}%
\bibitem [{\citenamefont {Stanescu}\ \emph {et~al.}(2009)\citenamefont
  {Stanescu}, \citenamefont {Galitski}, \citenamefont {Vaishnav}, \citenamefont
  {Clark},\ and\ \citenamefont {Das~Sarma}}]{stanescu:2009}%
  \BibitemOpen
  \bibfield  {author} {\bibinfo {author} {\bibfnamefont {T.~D.}\ \bibnamefont
  {Stanescu}}, \bibinfo {author} {\bibfnamefont {V.}~\bibnamefont {Galitski}},
  \bibinfo {author} {\bibfnamefont {J.~Y.}\ \bibnamefont {Vaishnav}}, \bibinfo
  {author} {\bibfnamefont {C.~W.}\ \bibnamefont {Clark}}, \ and\ \bibinfo
  {author} {\bibfnamefont {S.}~\bibnamefont {Das~Sarma}},\ }\href {\doibase
  10.1103/PhysRevA.79.053639} {\bibfield  {journal} {\bibinfo  {journal} {Phys.
  Rev. A}\ }\textbf {\bibinfo {volume} {79}},\ \bibinfo {pages} {053639}
  (\bibinfo {year} {2009})}\BibitemShut {NoStop}%
\bibitem [{\citenamefont {Goldman}\ \emph {et~al.}(2016)\citenamefont
  {Goldman}, \citenamefont {Budich},\ and\ \citenamefont
  {Zoller}}]{goldman:2016}%
  \BibitemOpen
  \bibfield  {author} {\bibinfo {author} {\bibfnamefont {N.}~\bibnamefont
  {Goldman}}, \bibinfo {author} {\bibfnamefont {J.~C.}\ \bibnamefont {Budich}},
  \ and\ \bibinfo {author} {\bibfnamefont {P.}~\bibnamefont {Zoller}},\
  }\href@noop {} {\bibfield  {journal} {\bibinfo  {journal} {Nature Physics}\
  }\textbf {\bibinfo {volume} {12}},\ \bibinfo {pages} {639} (\bibinfo {year}
  {2016})}\BibitemShut {NoStop}%
\bibitem [{\citenamefont {Rechtsman}\ \emph {et~al.}(2013)\citenamefont
  {Rechtsman}, \citenamefont {Zeuner}, \citenamefont {Plotnik}, \citenamefont
  {Lumer}, \citenamefont {Podolsky}, \citenamefont {Dreisow}, \citenamefont
  {Nolte}, \citenamefont {Segev},\ and\ \citenamefont
  {Szameit}}]{rechtsman:2013}%
  \BibitemOpen
  \bibfield  {author} {\bibinfo {author} {\bibfnamefont {M.~C.}\ \bibnamefont
  {Rechtsman}}, \bibinfo {author} {\bibfnamefont {J.~M.}\ \bibnamefont
  {Zeuner}}, \bibinfo {author} {\bibfnamefont {Y.}~\bibnamefont {Plotnik}},
  \bibinfo {author} {\bibfnamefont {Y.}~\bibnamefont {Lumer}}, \bibinfo
  {author} {\bibfnamefont {D.}~\bibnamefont {Podolsky}}, \bibinfo {author}
  {\bibfnamefont {F.}~\bibnamefont {Dreisow}}, \bibinfo {author} {\bibfnamefont
  {S.}~\bibnamefont {Nolte}}, \bibinfo {author} {\bibfnamefont
  {M.}~\bibnamefont {Segev}}, \ and\ \bibinfo {author} {\bibfnamefont
  {A.}~\bibnamefont {Szameit}},\ }\href@noop {} {\bibfield  {journal} {\bibinfo
   {journal} {Nature}\ }\textbf {\bibinfo {volume} {496}},\ \bibinfo {pages}
  {196} (\bibinfo {year} {2013})}\BibitemShut {NoStop}%
\bibitem [{\citenamefont {Lu}\ \emph {et~al.}(2014)\citenamefont {Lu},
  \citenamefont {Joannopoulos},\ and\ \citenamefont
  {Solja\v{c}i\'{c}}}]{lu:2014}%
  \BibitemOpen
  \bibfield  {author} {\bibinfo {author} {\bibfnamefont {L.}~\bibnamefont
  {Lu}}, \bibinfo {author} {\bibfnamefont {J.~D.}\ \bibnamefont
  {Joannopoulos}}, \ and\ \bibinfo {author} {\bibfnamefont {M.}~\bibnamefont
  {Solja\v{c}i\'{c}}},\ }\href@noop {} {\bibfield  {journal} {\bibinfo
  {journal} {Nature photonics}\ }\textbf {\bibinfo {volume} {8}},\ \bibinfo
  {pages} {821–829} (\bibinfo {year} {2014})}\BibitemShut {NoStop}%
\bibitem [{\citenamefont {Mittal}\ \emph {et~al.}(2014)\citenamefont {Mittal},
  \citenamefont {Fan}, \citenamefont {Faez}, \citenamefont {Migdall},
  \citenamefont {Taylor},\ and\ \citenamefont {Hafezi}}]{mittal:2014}%
  \BibitemOpen
  \bibfield  {author} {\bibinfo {author} {\bibfnamefont {S.}~\bibnamefont
  {Mittal}}, \bibinfo {author} {\bibfnamefont {J.}~\bibnamefont {Fan}},
  \bibinfo {author} {\bibfnamefont {S.}~\bibnamefont {Faez}}, \bibinfo {author}
  {\bibfnamefont {A.}~\bibnamefont {Migdall}}, \bibinfo {author} {\bibfnamefont
  {J.~M.}\ \bibnamefont {Taylor}}, \ and\ \bibinfo {author} {\bibfnamefont
  {M.}~\bibnamefont {Hafezi}},\ }\href {\doibase
  10.1103/PhysRevLett.113.087403} {\bibfield  {journal} {\bibinfo  {journal}
  {Phys. Rev. Lett.}\ }\textbf {\bibinfo {volume} {113}},\ \bibinfo {pages}
  {087403} (\bibinfo {year} {2014})}\BibitemShut {NoStop}%
\bibitem [{\citenamefont {Ningyuan}\ \emph {et~al.}(2015)\citenamefont
  {Ningyuan}, \citenamefont {Owens}, \citenamefont {Sommer}, \citenamefont
  {Schuster},\ and\ \citenamefont {Simon}}]{ningyuan:2015}%
  \BibitemOpen
  \bibfield  {author} {\bibinfo {author} {\bibfnamefont {J.}~\bibnamefont
  {Ningyuan}}, \bibinfo {author} {\bibfnamefont {C.}~\bibnamefont {Owens}},
  \bibinfo {author} {\bibfnamefont {A.}~\bibnamefont {Sommer}}, \bibinfo
  {author} {\bibfnamefont {D.}~\bibnamefont {Schuster}}, \ and\ \bibinfo
  {author} {\bibfnamefont {J.}~\bibnamefont {Simon}},\ }\href {\doibase
  10.1103/PhysRevX.5.021031} {\bibfield  {journal} {\bibinfo  {journal} {Phys.
  Rev. X}\ }\textbf {\bibinfo {volume} {5}},\ \bibinfo {pages} {021031}
  (\bibinfo {year} {2015})}\BibitemShut {NoStop}%
\bibitem [{\citenamefont {Mittal}\ \emph {et~al.}(2016)\citenamefont {Mittal},
  \citenamefont {Ganeshan}, \citenamefont {Fan}, \citenamefont {Vaezi},\ and\
  \citenamefont {Hafezi}}]{mittal:2016}%
  \BibitemOpen
  \bibfield  {author} {\bibinfo {author} {\bibfnamefont {S.}~\bibnamefont
  {Mittal}}, \bibinfo {author} {\bibfnamefont {S.}~\bibnamefont {Ganeshan}},
  \bibinfo {author} {\bibfnamefont {J.}~\bibnamefont {Fan}}, \bibinfo {author}
  {\bibfnamefont {A.}~\bibnamefont {Vaezi}}, \ and\ \bibinfo {author}
  {\bibfnamefont {M.}~\bibnamefont {Hafezi}},\ }\href@noop {} {\bibfield
  {journal} {\bibinfo  {journal} {Nature Physics}\ }\textbf {\bibinfo {volume}
  {10}},\ \bibinfo {pages} {180} (\bibinfo {year} {2016})}\BibitemShut
  {NoStop}%
\bibitem [{\citenamefont {Anderson}\ \emph {et~al.}(2016)\citenamefont
  {Anderson}, \citenamefont {Ma}, \citenamefont {Owens}, \citenamefont
  {Schuster},\ and\ \citenamefont {Simon}}]{anderson:2016}%
  \BibitemOpen
  \bibfield  {author} {\bibinfo {author} {\bibfnamefont {B.~M.}\ \bibnamefont
  {Anderson}}, \bibinfo {author} {\bibfnamefont {R.}~\bibnamefont {Ma}},
  \bibinfo {author} {\bibfnamefont {C.}~\bibnamefont {Owens}}, \bibinfo
  {author} {\bibfnamefont {D.~I.}\ \bibnamefont {Schuster}}, \ and\ \bibinfo
  {author} {\bibfnamefont {J.}~\bibnamefont {Simon}},\ }\href {\doibase
  10.1103/PhysRevX.6.041043} {\bibfield  {journal} {\bibinfo  {journal} {Phys.
  Rev. X}\ }\textbf {\bibinfo {volume} {6}},\ \bibinfo {pages} {041043}
  (\bibinfo {year} {2016})}\BibitemShut {NoStop}%
\bibitem [{\citenamefont {Cheng}\ \emph {et~al.}(2016)\citenamefont {Cheng},
  \citenamefont {Jouvaud}, \citenamefont {Ni}, \citenamefont {Mousavi},
  \citenamefont {Genack},\ and\ \citenamefont {Khanikaev}}]{cheng:2016}%
  \BibitemOpen
  \bibfield  {author} {\bibinfo {author} {\bibfnamefont {X.}~\bibnamefont
  {Cheng}}, \bibinfo {author} {\bibfnamefont {C.}~\bibnamefont {Jouvaud}},
  \bibinfo {author} {\bibfnamefont {X.}~\bibnamefont {Ni}}, \bibinfo {author}
  {\bibfnamefont {S.~H.}\ \bibnamefont {Mousavi}}, \bibinfo {author}
  {\bibfnamefont {A.~Z.}\ \bibnamefont {Genack}}, \ and\ \bibinfo {author}
  {\bibfnamefont {A.~B.}\ \bibnamefont {Khanikaev}},\ }\href@noop {} {\bibfield
   {journal} {\bibinfo  {journal} {Nature Materials}\ }\textbf {\bibinfo
  {volume} {15}},\ \bibinfo {pages} {542} (\bibinfo {year} {2016})}\BibitemShut
  {NoStop}%
\bibitem [{\citenamefont {Owens}\ \emph {et~al.}(2018)\citenamefont {Owens},
  \citenamefont {LaChapelle}, \citenamefont {Saxberg}, \citenamefont
  {Anderson}, \citenamefont {Ma}, \citenamefont {Simon},\ and\ \citenamefont
  {Schuster}}]{owens:2018}%
  \BibitemOpen
  \bibfield  {author} {\bibinfo {author} {\bibfnamefont {C.}~\bibnamefont
  {Owens}}, \bibinfo {author} {\bibfnamefont {A.}~\bibnamefont {LaChapelle}},
  \bibinfo {author} {\bibfnamefont {B.}~\bibnamefont {Saxberg}}, \bibinfo
  {author} {\bibfnamefont {B.~M.}\ \bibnamefont {Anderson}}, \bibinfo {author}
  {\bibfnamefont {R.}~\bibnamefont {Ma}}, \bibinfo {author} {\bibfnamefont
  {J.}~\bibnamefont {Simon}}, \ and\ \bibinfo {author} {\bibfnamefont {D.~I.}\
  \bibnamefont {Schuster}},\ }\href {\doibase 10.1103/PhysRevA.97.013818}
  {\bibfield  {journal} {\bibinfo  {journal} {Phys. Rev. A}\ }\textbf {\bibinfo
  {volume} {97}},\ \bibinfo {pages} {013818} (\bibinfo {year}
  {2018})}\BibitemShut {NoStop}%
\bibitem [{\citenamefont {Hadad}\ \emph {et~al.}(2018)\citenamefont {Hadad}, \citenamefont {Soric}, \citenamefont {Khanikaev}, \citenamefont {Al\`u}}]{hadad:2018}%
  \BibitemOpen
  \bibfield  {author} {\bibinfo {author} {\bibfnamefont {Y.}\ \bibnamefont
  {Hadad}},\ \bibinfo {author} {\bibfnamefont {J.C.}\ \bibnamefont
  {Soric}},\ \bibinfo {author} {\bibfnamefont {A.B.}\ \bibnamefont
  {Khanikaev},\ and\ \bibinfo {author} {\bibfnamefont {A.}\ \bibnamefont
  {Al\`u},\ }}}\href@noop {} {\bibfield  {journal} {\bibinfo  {journal}
  {Nature Electronics}\ }\textbf {\bibinfo {volume} {1}},\ \bibinfo {pages} {178}
  (\bibinfo {year} {2018})}\BibitemShut {NoStop}%
\bibitem [{\citenamefont {Kane}\ and\ \citenamefont
  {Lubensky}(2014)}]{kane:2014}%
  \BibitemOpen
  \bibfield  {author} {\bibinfo {author} {\bibfnamefont {C.~L.}\ \bibnamefont
  {Kane}}\ and\ \bibinfo {author} {\bibfnamefont {T.~C.}\ \bibnamefont
  {Lubensky}},\ }\href@noop {} {\bibfield  {journal} {\bibinfo  {journal}
  {Nature Physics}\ }\textbf {\bibinfo {volume} {10}},\ \bibinfo {pages} {39}
  (\bibinfo {year} {2014})}\BibitemShut {NoStop}%
\bibitem [{\citenamefont {Paulose}\ \emph {et~al.}(2015)\citenamefont
  {Paulose}, \citenamefont {Chen},\ and\ \citenamefont
  {Vitelli}}]{paulose:2015}%
  \BibitemOpen
  \bibfield  {author} {\bibinfo {author} {\bibfnamefont {J.}~\bibnamefont
  {Paulose}}, \bibinfo {author} {\bibfnamefont {B.~G.}\ \bibnamefont {Chen}}, \
  and\ \bibinfo {author} {\bibfnamefont {V.}~\bibnamefont {Vitelli}},\
  }\href@noop {} {\bibfield  {journal} {\bibinfo  {journal} {Nature Physics}\
  }\textbf {\bibinfo {volume} {11}},\ \bibinfo {pages} {153} (\bibinfo {year}
  {2015})}\BibitemShut {NoStop}%
\bibitem [{\citenamefont {S{\"u}sstrunk}\ and\ \citenamefont
  {Huber}(2016)}]{susstrunk:2016}%
  \BibitemOpen
  \bibfield  {author} {\bibinfo {author} {\bibfnamefont {R.}~\bibnamefont
  {S{\"u}sstrunk}}\ and\ \bibinfo {author} {\bibfnamefont {S.~D.}\ \bibnamefont
  {Huber}},\ }\href@noop {} {\bibfield  {journal} {\bibinfo  {journal}
  {Proceedings of the National Academy of Sciences}\ }\textbf {\bibinfo
  {volume} {113}},\ \bibinfo {pages} {E4767} (\bibinfo {year}
  {2016})}\BibitemShut {NoStop}%
\bibitem [{\citenamefont {Huber}(2016)}]{huber:2016}%
  \BibitemOpen
  \bibfield  {author} {\bibinfo {author} {\bibfnamefont {S.~D.}\ \bibnamefont
  {Huber}},\ }\href@noop {} {\bibfield  {journal} {\bibinfo  {journal} {Nature
  Physics}\ }\textbf {\bibinfo {volume} {12}},\ \bibinfo {pages} {621}
  (\bibinfo {year} {2016})}\BibitemShut {NoStop}%
\bibitem [{\citenamefont {Rudner}\ and\ \citenamefont
  {Levitov}(2009)}]{rudner:2009}%
  \BibitemOpen
  \bibfield  {author} {\bibinfo {author} {\bibfnamefont {M.~S.}\ \bibnamefont
  {Rudner}}\ and\ \bibinfo {author} {\bibfnamefont {L.~S.}\ \bibnamefont
  {Levitov}},\ }\href@noop {} {\bibfield  {journal} {\bibinfo  {journal} {Phys.
  Rev. Lett.}\ }\textbf {\bibinfo {volume} {102}},\ \bibinfo {pages} {065703}
  (\bibinfo {year} {2009})}\BibitemShut {NoStop}%
\bibitem [{\citenamefont {Diehl}\ \emph {et~al.}(2011)\citenamefont {Diehl},
  \citenamefont {Rico}, \citenamefont {Baranov},\ and\ \citenamefont
  {Zoller}}]{diehl:2011}%
  \BibitemOpen
  \bibfield  {author} {\bibinfo {author} {\bibfnamefont {S.}~\bibnamefont
  {Diehl}}, \bibinfo {author} {\bibfnamefont {E.}~\bibnamefont {Rico}},
  \bibinfo {author} {\bibfnamefont {M.~A.}\ \bibnamefont {Baranov}}, \ and\
  \bibinfo {author} {\bibfnamefont {P.}~\bibnamefont {Zoller}},\ }\href@noop {}
  {\bibfield  {journal} {\bibinfo  {journal} {Nature Physics}\ }\textbf
  {\bibinfo {volume} {7}},\ \bibinfo {pages} {971} (\bibinfo {year}
  {2011})}\BibitemShut {NoStop}%
\bibitem [{\citenamefont {Zeuner}\ \emph {et~al.}(2015)\citenamefont {Zeuner},
  \citenamefont {Rechtsman}, \citenamefont {Plotnik}, \citenamefont {Lumer},
  \citenamefont {Nolte}, \citenamefont {Rudner}, \citenamefont {Segev},\ and\
  \citenamefont {Szameit}}]{zeuner:2015}%
  \BibitemOpen
  \bibfield  {author} {\bibinfo {author} {\bibfnamefont {J.~M.}\ \bibnamefont
  {Zeuner}}, \bibinfo {author} {\bibfnamefont {M.~C.}\ \bibnamefont
  {Rechtsman}}, \bibinfo {author} {\bibfnamefont {Y.}~\bibnamefont {Plotnik}},
  \bibinfo {author} {\bibfnamefont {Y.}~\bibnamefont {Lumer}}, \bibinfo
  {author} {\bibfnamefont {S.}~\bibnamefont {Nolte}}, \bibinfo {author}
  {\bibfnamefont {M.~S.}\ \bibnamefont {Rudner}}, \bibinfo {author}
  {\bibfnamefont {M.}~\bibnamefont {Segev}}, \ and\ \bibinfo {author}
  {\bibfnamefont {A.}~\bibnamefont {Szameit}},\ }\href {\doibase
  10.1103/PhysRevLett.115.040402} {\bibfield  {journal} {\bibinfo  {journal}
  {Phys. Rev. Lett.}\ }\textbf {\bibinfo {volume} {115}},\ \bibinfo {pages}
  {040402} (\bibinfo {year} {2015})}\BibitemShut {NoStop}%
\bibitem [{\citenamefont {Rudner}\ \emph {et~al.}(2016)\citenamefont {Rudner},
  \citenamefont {Levin},\ and\ \citenamefont {Levitov}}]{rudner:2016}%
  \BibitemOpen
  \bibfield  {author} {\bibinfo {author} {\bibfnamefont {M.~S.}\ \bibnamefont
  {Rudner}}, \bibinfo {author} {\bibfnamefont {M.}~\bibnamefont {Levin}}, \
  and\ \bibinfo {author} {\bibfnamefont {L.~S.}\ \bibnamefont {Levitov}},\
  }\href@noop {} {\bibfield  {journal} {\bibinfo  {journal} {arXiv preprint
  arXiv:1605.07652}\ } (\bibinfo {year} {2016})}\BibitemShut {NoStop}%
\bibitem [{\citenamefont {Leykam}\ \emph {et~al.}(2017)\citenamefont {Leykam},
  \citenamefont {Bliokh}, \citenamefont {Huang}, \citenamefont {Chong},\ and\
  \citenamefont {Nori}}]{leykam:2017}%
  \BibitemOpen
  \bibfield  {author} {\bibinfo {author} {\bibfnamefont {D.}~\bibnamefont
  {Leykam}}, \bibinfo {author} {\bibfnamefont {K.~Y.}\ \bibnamefont {Bliokh}},
  \bibinfo {author} {\bibfnamefont {C.}~\bibnamefont {Huang}}, \bibinfo
  {author} {\bibfnamefont {Y.~D.}\ \bibnamefont {Chong}}, \ and\ \bibinfo
  {author} {\bibfnamefont {F.}~\bibnamefont {Nori}},\ }\href {\doibase
  10.1103/PhysRevLett.118.040401} {\bibfield  {journal} {\bibinfo  {journal}
  {Phys. Rev. Lett.}\ }\textbf {\bibinfo {volume} {118}},\ \bibinfo {pages}
  {040401} (\bibinfo {year} {2017})}\BibitemShut {NoStop}%
\bibitem [{\citenamefont {Rakovszky}\ \emph {et~al.}(2017)\citenamefont
  {Rakovszky}, \citenamefont {Asb\'oth},\ and\ \citenamefont
  {Alberti}}]{rakovszky:2017}%
  \BibitemOpen
  \bibfield  {author} {\bibinfo {author} {\bibfnamefont {T.}~\bibnamefont
  {Rakovszky}}, \bibinfo {author} {\bibfnamefont {J.~K.}\ \bibnamefont
  {Asb\'oth}}, \ and\ \bibinfo {author} {\bibfnamefont {A.}~\bibnamefont
  {Alberti}},\ }\href {\doibase 10.1103/PhysRevB.95.201407} {\bibfield
  {journal} {\bibinfo  {journal} {Phys. Rev. B}\ }\textbf {\bibinfo {volume}
  {95}},\ \bibinfo {pages} {201407} (\bibinfo {year} {2017})}\BibitemShut
  {NoStop}%
\bibitem [{\citenamefont {Zhan}\ \emph {et~al.}(2017)\citenamefont {Zhan},
  \citenamefont {Xiao}, \citenamefont {Bian}, \citenamefont {Wang},
  \citenamefont {Qiu}, \citenamefont {Sanders}, \citenamefont {Yi},\ and\
  \citenamefont {Xue}}]{zhan:2017}%
  \BibitemOpen
  \bibfield  {author} {\bibinfo {author} {\bibfnamefont {X.}~\bibnamefont
  {Zhan}}, \bibinfo {author} {\bibfnamefont {L.}~\bibnamefont {Xiao}}, \bibinfo
  {author} {\bibfnamefont {Z.}~\bibnamefont {Bian}}, \bibinfo {author}
  {\bibfnamefont {K.}~\bibnamefont {Wang}}, \bibinfo {author} {\bibfnamefont
  {X.}~\bibnamefont {Qiu}}, \bibinfo {author} {\bibfnamefont {B.~C.}\
  \bibnamefont {Sanders}}, \bibinfo {author} {\bibfnamefont {W.}~\bibnamefont
  {Yi}}, \ and\ \bibinfo {author} {\bibfnamefont {P.}~\bibnamefont {Xue}},\
  }\href {\doibase 10.1103/PhysRevLett.119.130501} {\bibfield  {journal}
  {\bibinfo  {journal} {Phys. Rev. Lett.}\ }\textbf {\bibinfo {volume} {119}},\
  \bibinfo {pages} {130501} (\bibinfo {year} {2017})}\BibitemShut {NoStop}%
\bibitem [{\citenamefont {Weimann}\ \emph {et~al.}(2017)\citenamefont
  {Weimann}, \citenamefont {Kremer}, \citenamefont {Plotnik}, \citenamefont
  {Lumer}, \citenamefont {Nolte}, \citenamefont {Makris}, \citenamefont
  {Segev}, \citenamefont {Rechtsman},\ and\ \citenamefont
  {Szameitr}}]{weimann:2017}%
  \BibitemOpen
  \bibfield  {author} {\bibinfo {author} {\bibfnamefont {S.}~\bibnamefont
  {Weimann}}, \bibinfo {author} {\bibfnamefont {M.}~\bibnamefont {Kremer}},
  \bibinfo {author} {\bibfnamefont {Y.}~\bibnamefont {Plotnik}}, \bibinfo
  {author} {\bibfnamefont {Y.}~\bibnamefont {Lumer}}, \bibinfo {author}
  {\bibfnamefont {S.}~\bibnamefont {Nolte}}, \bibinfo {author} {\bibfnamefont
  {K.~G.}\ \bibnamefont {Makris}}, \bibinfo {author} {\bibfnamefont
  {M.}~\bibnamefont {Segev}}, \bibinfo {author} {\bibfnamefont {M.~C.}\
  \bibnamefont {Rechtsman}}, \ and\ \bibinfo {author} {\bibfnamefont
  {A.}~\bibnamefont {Szameitr}},\ }\href@noop {} {\bibfield  {journal}
  {\bibinfo  {journal} {Nature Materials}\ }\textbf {\bibinfo {volume} {16}},\
  \bibinfo {pages} {433–438} (\bibinfo {year} {2017})}\BibitemShut {NoStop}%
\bibitem [{Note1()}]{Note1}%
  \BibitemOpen
  \bibinfo {note} {In direct space, the dark state corresponds to voltages on
  neighboring $A$-sites oscillating in anti-phase ($k = \pi $), with zero
  amplitude on the lossy $B$-sites.}\BibitemShut {Stop}%
\bibitem [{\citenamefont {Dormand}\ and\ \citenamefont
  {Prince}(1980)}]{dormand:1980}%
  \BibitemOpen
  \bibfield  {author} {\bibinfo {author} {\bibfnamefont {J.~R.}\ \bibnamefont
  {Dormand}}\ and\ \bibinfo {author} {\bibfnamefont {P.~J.}\ \bibnamefont
  {Prince}},\ }\href@noop {} {\bibfield  {journal} {\bibinfo  {journal} {J.
  Comp. Appl. Math}\ }\textbf {\bibinfo {volume} {6}},\ \bibinfo {pages} {19}
  (\bibinfo {year} {1980})}\BibitemShut {NoStop}%
\bibitem [{\citenamefont {Shampine}\ and\ \citenamefont
  {Reichelt}(1997)}]{shampine:1997}%
  \BibitemOpen
  \bibfield  {author} {\bibinfo {author} {\bibfnamefont {L.~F.}\ \bibnamefont
  {Shampine}}\ and\ \bibinfo {author} {\bibfnamefont {M.~W.}\ \bibnamefont
  {Reichelt}},\ }\href@noop {} {\bibfield  {journal} {\bibinfo  {journal} {SIAM
  Journal on Scientific Computing}\ }\textbf {\bibinfo {volume} {18}},\
  \bibinfo {pages} {1} (\bibinfo {year} {1997})}\BibitemShut {NoStop}%
\bibitem [{Note3()}]{Note3}%
  \BibitemOpen
  \bibinfo {note} {Note that a finite-size lattice with periodic boundary
  conditions and an {\protect \it odd} number of unit cells, $N$, does not
  support a true dark state (Fig.~S2). In this case, the allowed set of crystal
  momentum values, $k_n = 2 \pi n / N$, where $n = 0, 1, 2, \protect \ldots ,
  N-1$, does not include $k = \pi $. Modes with $k$ values close to $\pi $
  still exhibit long lifetimes, with lifetime increasing when $\alpha $ is near
  1/2 and the number of unit cells increases.}\BibitemShut {Stop}%
\bibitem [{\citenamefont {Gong}\ \emph {et~al.}(2018)\citenamefont {Gong},
  \citenamefont {Ashida},\ \citenamefont {Kawabata} \citenamefont {Takasan},\ \citenamefont {Higashikawa},\ \citenamefont {Ueda}}]{gong:2018}%
  \BibitemOpen
  \bibfield  {author} {\bibinfo {author} {\bibfnamefont {Z.}\ \bibnamefont
  {Gong}}, \bibinfo {author} {\bibfnamefont {Y.}~\bibnamefont {Ashida}}, \ \bibinfo {author} {\bibfnamefont {K.}\ \bibnamefont {Kawabata}}, \bibinfo {author} {\bibfnamefont {K.}\ \bibnamefont {Takasan}}, \ \bibinfo {author} {\bibfnamefont {S.}\ \bibnamefont
  {Higashikawa}}, \ and\ \bibinfo {author} {\bibfnamefont {M.}\ \bibnamefont
  {Ueda}},
  }\href@noop {} {\bibfield  {journal} {\bibinfo  {journal} {arXiv preprint
  arXiv:1802.07964}\ } (\bibinfo {year} {2018})}\BibitemShut {NoStop}%
\end{thebibliography}
\end{document}